\documentclass[12pt,preprint]{aastex}
\usepackage{amsmath}

\begin{document}

\title{An Illustration of Modeling Cataclysmic Variables: HST, FUSE, 
SDSS Spectra of SDSSJ080908.39+381406.2\footnotemark[1]}
\footnotetext[1]
{Based on observations made with the NASA/ESA Hubble Space Telescope, obtained at the
Space Telescope Science Institute, which is operated by the Association of Universities 
for Research in Astronomy, Inc. under NASA contract NAS5-26555, the NASA-CNES-CSA
{\it Far Ultraviolet Explorer}, which is operated for NASA by the Johns Hopkins University
under NASA contract NAS5-32985, and with the Apache Point 
Observatory 3.5 m telescope, which is operated by the Astrophysical Research Corporation.}


\author{Albert P. Linnell$^2$, D.W.Hoard$^3$, Paula Szkody$^4$, Knox S. Long$^5$,
Ivan Hubeny$^6$,Boris G\"{a}nsicke$^7$, and Edward M.Sion$^8$}
\affil{$^2$Department of Astronomy, University of Washington, Box 351580, Seattle,
WA 98195-1580\\
$^3$Spitzer Science Center, California Institute of Technology, Mail Code 220-6,
1200 E. California Blvd., Pasadena, CA 91125\\
$^4$Department of Astronomy, University of Washington, Box 351580, Seattle, 
WA 98195-1580\\
$^5$Space Telescope Science Institute, 3700 San Martin Drive, Baltimore, MD 21218\\
$^6$Steward Observatory and Department of Astronomy,
University of Arizona, Tucson, AZ 85721\\
$^7$Department of Physics, University of Warwick, CV4 7AL,
Coventry, UK\\
$^8$Department of Astronomy and Astrophysics, Villanova University,
Villanova, PA 19085\\}

\email{$^2$linnell@astro.washington.edu\\
$^3$hoard@ipac.caltech.edu\\
$^4$szkody@astro.washington.edu\\
$^5$long@stsci.edu\\
$^6$hubeny@as.arizona.edu\\
$^7$boris.gaensicke@warwick.ac.uk\\
$^8$edward.sion@villanova.edu\\
}

\begin{abstract}

{\it FUSE}, {\it HST} and SDSS spectra of the cataclysmic variable SDSSJ080908.39
+381406.2 provide a spectral flux distribution from 900--9200\AA.
This data set is used to illustrate procedures for calculating and
testing system models. The spectra are not contemporaneous.
The illustrations are based on a system with a $1.0M_{\odot}$ white dwarf,	a $0.30M_{\odot}$,
3500K, Roche lobe-filling secondary star, and an accretion disk extending to the
tidal cutoff radius. Assuming a similar accretion state for the non-simultaneous spectra,
the best standard model fit
is with a
mass transfer rate of $3.0{\times}10^{-9}M_{\odot} {\rm yr^{-1}}$.
Extensive
simulations demonstrate that the accretion disk must be truncated at its inner edge 
if the temperature
profile follows the standard model, but truncated models face severe objections, which we 
address. Following additional simulation tests, we obtain a model accretion disk with a temperature 
profile comparable to
the profile for SW Sex as determined from tomographic image reconstruction.
This model fits the discovery SDSS spectrum well but has a flux deficit in the UV and FUV.
Emission from a white dwarf is a plausable source of additional flux. 
Adding this source to the disk synthetic spectrum
produces FUV flux that can explain the observed flux. 

An additional (archival) SDSS spectrum is fainter by about 0.3 magnitude in the optical.
Additional analysis showed that UV residuals from a model fitting the archival optical
wavelength spectrum are
unacceptably large. Contemporaneous spectra from all wavelength regions
would be necessary for a reliable system model. 
Our discussion illustrates how this conclusion follows from the system models.

\end{abstract}



\section{Introduction}
\label{s-intro}

Cataclysmic variables (CVs) are semi-detached binary stars in 
which a late main sequence star loses mass onto a white dwarf (WD) 
via Roche lobe overflow.  
In systems containing a non-magnetic WD, 
accretion proceeds through a viscous disk.  However, in systems 
containing a magnetic WD, accretion may proceed directly from the 
inner Lagrangian point onto field lines in the strong field case, 
or through a partial disk in the case of intermediate field 
strength.  
The group of CVs that have typical orbital periods between 3-4 hrs, 
strong single-peaked Balmer emission,
strong HeII emission and deep central absorption in the Balmer and
HeI lines (usually near orbital phase 0.5) are called the SW Sex
stars (e.g., \citet{honey86,szkody90,thor91}). As a group, their orbital
periods place the SW Sex stars at the top of the period gap, which likely is
an aspect of the normal secular evolution of the CV population \citep{warner95}.
The SW Sex stars typically have high mass accretion
rates and a few of them may have intermediate strength magnetic fields as evinced
from weak polarization \citep{r2001}.
See \citet{warner95} for a thorough review of CV types and behaviors.

The observational data that are commonly available to study CVs include
spectra and photometry. Deduction of the physical properties of these binary
stars depends on simulation of the observations, based on a physical model
believed to represent reality with adequate accuracy.
The accretion disk is typically the most important radiating component in a CV over
a wide range of wavelengths.
A common practice has been to use synthetic stellar spectra to
simulate the accretion disk (\citet{lad87,lad94}). More recently, \citet{wh98}
have calculated a set of models based on the TLUSTY program.
It is desirable to have a simulation program that calculates an
accretion disk {\it ab initio}, given an orbital inclination,
WD mass, and mass transfer rate. A combination of the
programs TLUSTY \citep{h88} and BINSYN \citep{linnell96} provides this feature.
The details of how the calculation takes place, in the context of a specific
system, have not been described in the literature. 
The availability of both $HST$ and $FUSE$ data combined with a good SDSS spectrum
of a bright system (few systems provide this combination) 
provides the impetus for modeling this system. Since the data were not contemporaneous,
a definitive solution is not available. However, as we show later, we are
able to find a solution which satisfactorily represents the data set.
At the same time, the
data set provides a very useful basis to illustrate the details of building and
testing system models, and the results turn out to be of considerable
intrinsic interest.

\section{The SDSSJ0809 system}

The Sloan Digital Sky Survey (York et al. 2000)
provides
5 color photometry (ugriz) and spectra (3800-9200\AA\ at a resolution
of 3\AA) of the survey objects.
The survey has identified a large number of CV systems (Szkody et al. 2002,
2003, 2004, 2005). 
 
During the survey, SDSS J080908.39+381406.2 (hereafter abbreviated as
SDSSJ0809) was discovered as a bright CV (g$\sim$15.6)
(Szkody et al. 2003).   
The discovery paper provides the photometric colors and optical
spectrum of SDSSJ0809 as well as the results of time-resolved
followup spectra obtained over the course of 3 hrs. The SDSS spectrum
shows a steep blue continuum with strong Balmer emission lines as
well as prominent emission from \ion{He}{2} $\lambda4686$ and the CN blend at 4640\AA. 
The time-resolved spectra show deep and transient absorption in the Balmer and HeI lines.
If the H${\alpha}$ and H${\beta}$ radial velocities indicate orbital phases, then
the deep absorptions occur
near orbital phase 0.5 (Szkody et al. 2003), 
which is a hallmark of the SW Sex stars. However, we lack any confirming evidence (eclipses) and
we are presently unable to make any definite statement about orbital phases.
There
was no noticeable eclipse during the 3 hrs of observations and the radial velocity
curve constructed from the H$\alpha$ and H$\beta$ emission lines
indicated an orbital period near 2.4 hrs. However, the shortness of
the data interval precluded the determination of an accurate period. Recent photometry
over many nights indicates that the period is 192 min (more
typical for SW Sex stars) and
there may be a shallow or partial eclipse (Rodriguez-Gill et al., in preparation) with 
amplitude of less than
about 0.1 magnitude. We will use the 192 min period throughout this paper.

Due to the brightness of this system, ultraviolet measurements were
possible with both the Far Ultraviolet Spectroscopic Explorer ({\it FUSE})
 and the Hubble Space Telescope ({\it HST}) and were used to explore the
accretion parameters and geometry of the high accretion regime. In
the high inclination SW Sex star DW UMa, which shows deep eclipses,
 the accretion disk completely obscures the WD and inner
disk during high states of accretion (Knigge et al. 2004). When
the accretion is reduced during low states of mass transfer, the
underlying WD is revealed and it is very hot (near 50,000K \citep{ab2003}).
Most SW Sex stars have high inclinations and deep eclipses such 
that, as in DW UMa, the WD is hidden from view by the accretion disk.  
We observed SDSSJ0809 in the hope that its apparent lower inclination 
(as evinced by the lack of deep eclipses) might permit a more direct view of 
the WD and inner disk, and thus allow us to explore the characteristics 
of these components of an SW Sex star.
An objective of this investigation is to determine
the extent to which the
large SDSS spectral range, combined with spectra from $HST$ and $FUSE$, together with
related collateral information,
permit development of a self-consistent model.
 
As the following sections indicate, we have a single $HST$ spectrum, and two
sets of SDSS and $FUSE$ observations. None of the observations are simultaneous.
The photometric observations Rodriguez-Gill et al. were obtained some 300 days after the $HST$
exposure and continued over an interval of 26 days. The photometry shows only minor orbital
variation, with an amplitude of 0.1 magnitude that may indicate a grazing eclipse, but
with no systematic trend. There is consistent
overlap of the $HST$ and one of the $FUSE$ observations. Since the SDSS photometry,
the first SDSS spectrum, the subsequent APO spectroscopy, the Rodriguez-Gil photometry and the
optical measurement from the $HST$ snapshot all agree (to $\pm$0.1 mag), 
we first treated the system as if it is constant
at all observed wavelengths longward of the $FUSE$ observations.

\section{{\it FUSE} Observations and Data Processing}
\label{s-mainobs}

\begin{figure}[tb]
\epsscale{0.97}
\plotone{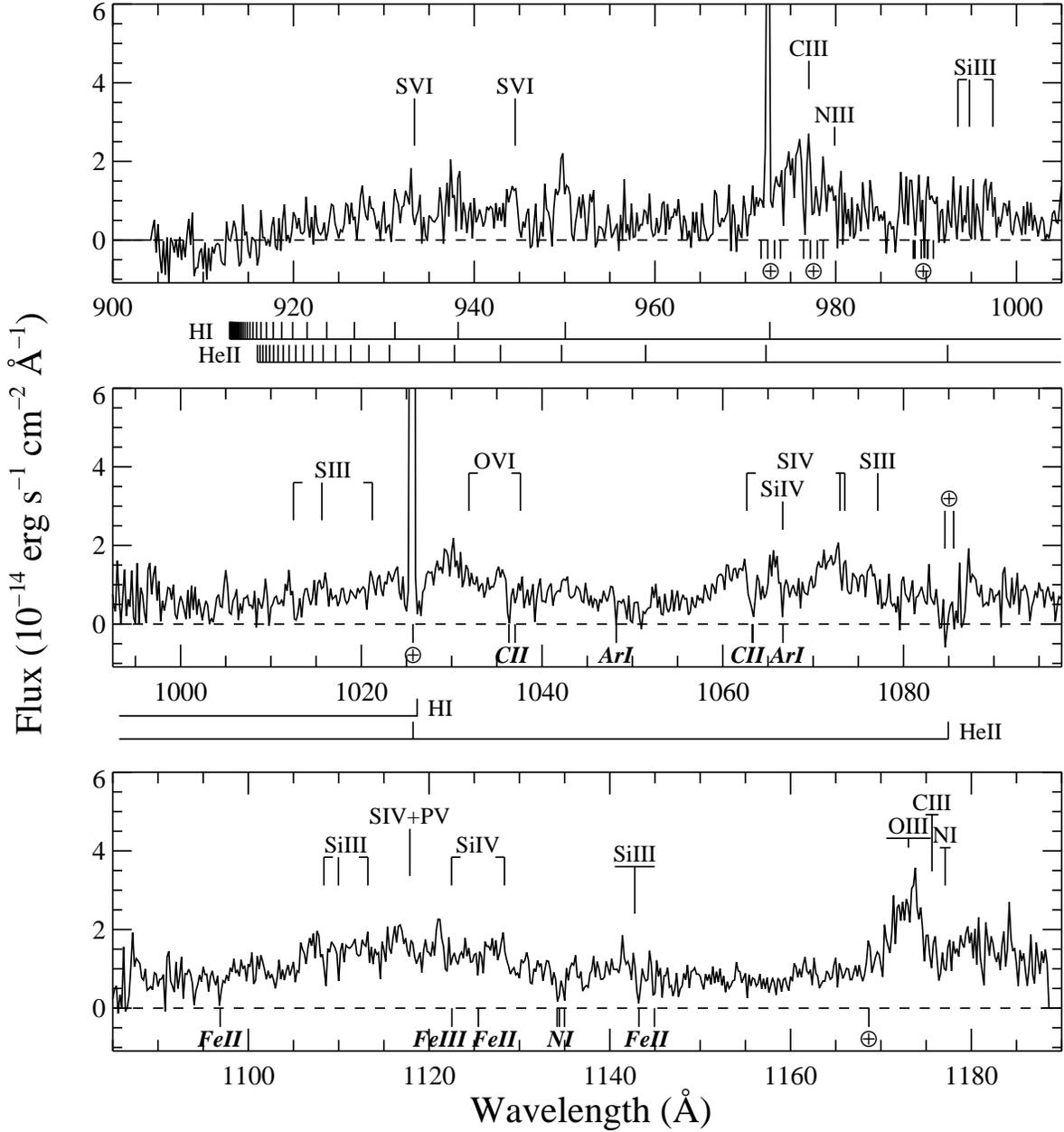}
\epsscale{1.00}
\figcaption{Average FUV spectrum of SDSSJ0809 from Observation 1 (2003), binned to a 
dispersion of 0.20 \AA\ pixel$^{-1}$.  The spectrum spans 
915--1185 \AA; the top and bottom panels overlap the ends of the 
middle panel by 7.5 \AA.  Prominent features often found
in FUV spectra of CVs are indicated, although not all of them are
present in SDSSJ0809 (see text).  
Airglow lines ($\oplus$; identified from \citealt{feldman01}) 
and probable ISM lines (ions in italics) are indicated 
below the spectrum.  
The airglow lines have been truncated at the upper flux limit 
of the plot.  Wavelengths of \ion{H}{1} and \ion{He}{2} 
transitions are shown between the panels.
\label{f-fuvspec1}}
\end{figure}

\begin{figure}[tb]
\epsscale{0.97}
\plotone{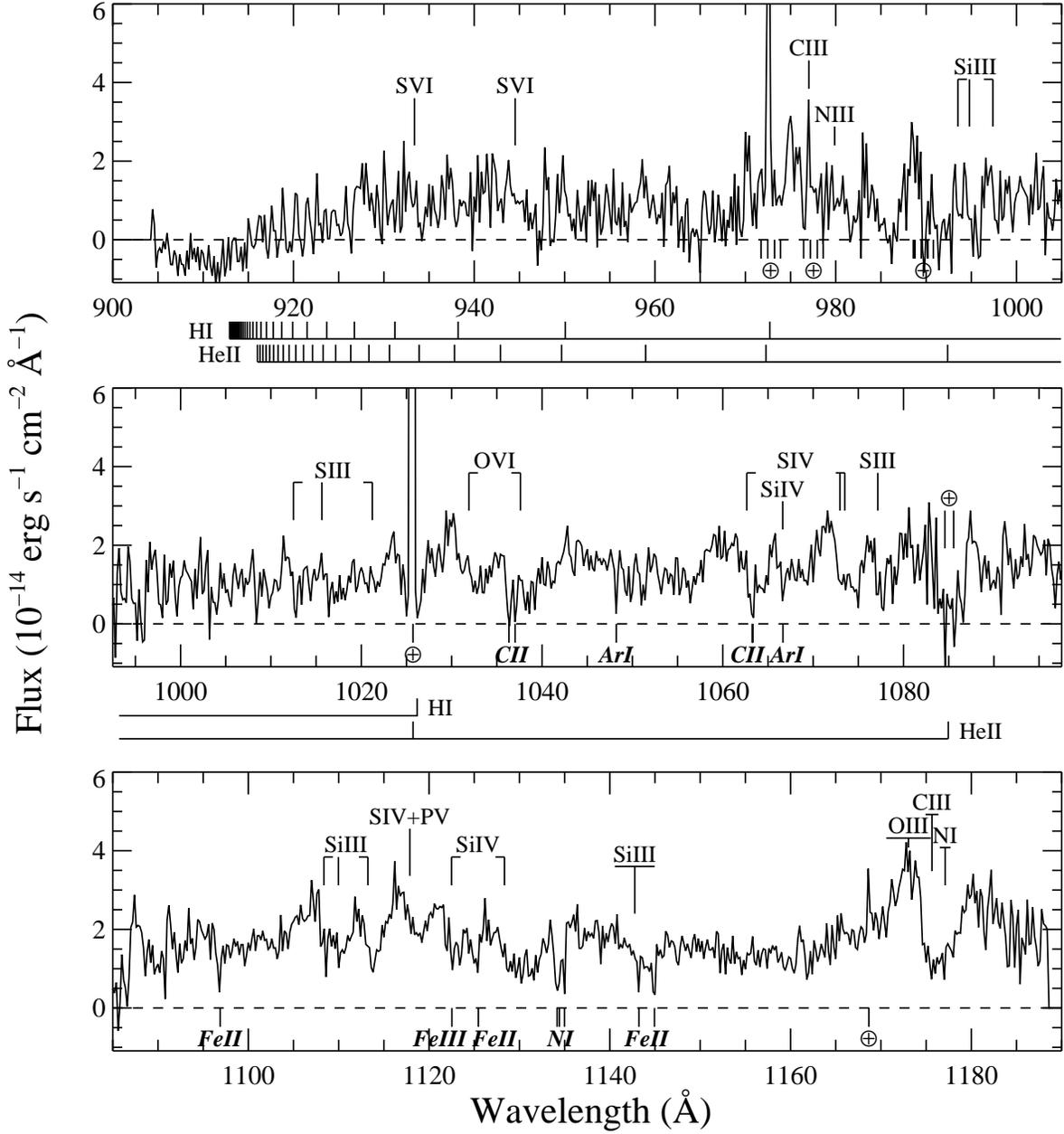}
\epsscale{1.00}
\figcaption{As in Figure \ref{f-fuvspec1} but showing the average FUV spectrum of 
SDSSJ0809 from Observation 2 (2004).
\label{f2}}
\end{figure}

We observed SDSSJ0809 with {\it FUSE} during two time intervals approximately 1 year apart; 
Observation 1 occurred on 28 Mar 2003 and Observation 2 occurred on 16--17 Mar 2004.  
Because of the {\it FUSE} orbital constraints, each observation was divided into several 
exposures (4 exposures for Observation 1, 
8 exposures for Observation 2) -- see Table \ref{t-fuselog} 
for the {\it FUSE} observing log.
All data were obtained using the LWRS aperture and TTAG accumulation mode 
(for {\it FUSE} spacecraft and instrument details see, for example, 
\citealt{sahnow00}\footnote{Also see the {\it FUSE} Science Center 
web page at \url{http://fuse.pha.jhu.edu/}.}).  We used the 
CalFuse v3.0.7 pipeline software to prepare intermediate data files 
from the raw data files obtained during each {\it FUSE} exposure.  
We then used the IDL routine cf\_edit v2.9\footnote{Written by Don 
Lindler and available at 
\url{http://fuse.pha.jhu.edu/analysis/fuse\_idl\_tools.html}.}, as well 
as custom-built IDL routines, to extract spectra from the various 
detector and mirror segments and combine them into a time-averaged 
spectrum for each observation.  The time-averaged spectra have total equivalent 
exposure times of $\approx 8.4$ ksec for Observation 1 and $\approx 10$ ksec for Observation 2 
(total exposure times from the 
mirror/detector pairs differ by up to $\sim5$\% due to differences in 
rejected data; the combined spectrum accounts for this by weighting by 
exposure time when combining data from different mirror/detector pairs).  
The final combined spectra (see Figures \ref{f-fuvspec1} and \ref{f2}) were rebinned 
onto a uniform wavelength scale with dispersion 0.20 \AA\ pixel$^{-1}$ 
by averaging flux points from the original dispersion 
(0.013 \AA\ pixel$^{-1}$) into wavelength bins of width 0.20 \AA.

\section{Analysis of FUV Spectra}

Our initial plan was to combine the two {\it FUSE} data sets to improve 
the S/N of the final time-averaged spectrum.  However, it was apparent 
that the spectra obtained from the two observations were somewhat 
different, so we treated them separately.  Shortward of 
$\approx1000$\AA\, the two spectra are approximately equivalent, 
although this apparent agreement is likely due mainly to the weak 
detection at these short wavelengths.  None of the sulfur or silicon 
emission lines often seen in FUV spectra of CVs in this wavelength 
region (e.g., \citealt{ho2003, froning01, long94}) are detected in 
SDSSJ0809 (see the top panels of Figures \ref{f-fuvspec1} and 
\ref{f2}).  The \ion{C}{3}, \ion{N}{3} emission complex is 
possibly present in both spectra; however, it coincides with the 
wavelengths of a series of airglow features, which makes a definitive 
detection difficult.

At longer wavelengths, the Observation 2 spectrum is approximately 
twice as bright in the continuum as the Observation 1 spectrum; for 
example, the ``mean'' continuum level at 1155\AA\ is 
$\approx0.75\times10^{-14}$ erg s$^{-1}$ cm$^{-2}$ \AA$^{-1}$ 
in the Observation 1 spectrum and 
$\approx1.5\times10^{-14}$ erg s$^{-1}$ cm$^{-2}$ \AA$^{-1}$ 
in the Observation 2 spectrum.

There are also differences in the ionic features present 
at $\lambda > 1000$\AA\ in each spectrum.  In the Observation 1 
spectrum, the S+Si line complexes at 1055--1085\AA\ and 1105--1130\AA\ 
are present as broad, undifferentiated ``humps'' in the continuum.  
However, in the Observation 2 spectrum, these same wavelength regions 
contain clearly differentiated emission and/or absorption features.  
The most striking difference is in the C+N+O line complex at 
1165--1185\AA.  In the Observation 1 spectrum, this is present mainly 
as a fairly strong emission feature (strongest of the non-airglow lines) 
at the wavelengths spanned by the \ion{O}{3} multiplet.  In the 
Observation 2 spectrum, on the other hand, this line complex displays 
a profile consisting of a broad hump from 1165--1185\AA, with a 
possible discrete \ion{O}{3} emission component, and deep absorption 
(reaching down to at least the level of the underlying continuum) 
at the wavelengths spanned by the \ion{C}{3}+\ion{N}{1} multiplets. 

The \ion{O}{6} lines at $\lambda\lambda1031.9, 1037.6$\AA, which are 
often among the most prominent emission features in the FUV spectra of 
CVs with ongoing accretion (e.g., 
\citealt{h2005, ho2003, hoard02, mauche99}), are only weakly present 
in both spectra of SDSSJ0809.  In addition, the line profiles of 
\ion{O}{6} appear to be in emission on the blue side and absorbed on 
the red side (similar to the shape of the C+N+O line complex at 
1165--1185\AA~in Observation 2).  Detection of the \ion{S}{3} multiplet 
at 1012--1021\AA\ is even weaker (or non-existent) compared to 
\ion{O}{6} in both spectra.  Neither spectrum displays detectable 
\ion{Si}{3} features at 1144\AA.

The total exposure time for each of the time-averaged FUV spectra 
of SDSSJ0809 is shorter than the orbital period of the CV (75--85\% 
of $P_{\rm orb}=192$ min). However, the total orbital phase coverage 
during a given observation differs from the other observation by 
only about 10\% and spans most of an orbit.  Thus, we believe that 
the differences between the spectra are not likely to 
be due only to differences in lines-of-sight through the orbital 
geometry.  In any case, the current estimate of the orbital period 
of SDSSJ0809 is not known precisely enough to preserve cycle count 
well enough from Observation 1 to Observation 2 that we could 
compare relative phases of the two {\it FUSE} spectra. 

The Observation 1 spectrum, which we designate as $FUSE1$, 
accords with the HST spectrum in the wavelength overlap region better 
than the Observation 2 spectrum. 
The Observation 2 spectrum difference from the 
Observation 1 spectrum could arise from a change in the accretion 
disk (discussed subsequently) in the time interval between the two 
observations.

\section{{\it HST} Observations}

SDSSJ0809 was observed on 2004 Apr 29 as part of an {\it HST} 
Space Telescope Imaging Spectrograph (STIS) snapshot program 
(see \citet{gaen} for a description of this program). Using 
the STIS/CCD acquisition image, we determined that the 
magnitude in the F28x50LP filter ($\sim5500-11,000$\AA) was 
about 15.5.  The STIS spectrum is an 800-sec exposure obtained 
with the G140L grating, in accum mode, which provided wavelength coverage 
from 1100--1700\AA\ with a resolution of 1.2\AA. 
The data were reduced with CALSTIS (V2.13b) which corrected 
for the decaying sensitivity of the grating.
 
The spectrum is shown in Figure 3. The
continuum displays a downturn near Ly$\alpha$ and weak emission lines
of CIII, NV, CII, SiIV and CIV can be identified. The lines are
noticeably weaker 
than those in the high state STIS spectrum of DW UMa (Knigge et al. 2004), 
consistent with a lower inclination for SDSSJ0809. 
The absorption cores at the
SiIV doublet wavelengths, superposed on the overlapping doublet emission, 
and the doubled appearance of the CIV lines are similar to the
STIS spectrum of DW UMa at phase 0.5 (Knigge et al. 2004), when
the absorption events characteristic of SW Sex stars are most
prominent.
A number of additional line identifications in Figure 3 mark 
absorption cores associated with broad emission features; 
in most cases the absorption core goes below the continuum level, 
indicating a substantial column density along the line of sight.  
Unfortunately, we cannot determine the phase of the SDSSJ0809 STIS spectrum 
as there currently is no orbital ephemeris.

\section{SDSS  and APO Observations}

The discovery spectrum of SDSS0809 
from MJD52251 is described in \citet*{sz03} 
and is reproduced here in Figure 4.
The
exposure time, 5803 sec, spans approximately half an orbital cycle.
There is no indication of a Balmer discontinuity.
The SDSS photometry for SDSSJ0809 gave g=15.61, and the 5500\AA~flux in the SDSS spectrum
gave an approximate $V$ magnitude of $\sim15.4$.

The Apache Point Observatory (APO) follow-up spectra, obtained over 2.9 hours, provided a radial 
velocity curve in both H$\alpha$ and H$\beta$ \citep*{sz03}. 
The time-resolved APO spectra
showed the strong modulation and deep transient absorption in the He I and Balmer line cores that
characterize SW Sex systems \citep*{sz03}. He II and CN emission are present 
at all phases of 
the APO spectra.

\section{CV accretion disks}
 
\citet{fkr92} discuss the theory of accretion disks. They show that the effective
temperature of an accretion disk, in hydrostatic equilibrium and timewise invariant,
at a distance $R$ from the accretion disk center, is given by the equation

\begin{equation}
T_{\rm eff}(R)=\Bigl\{{\frac{3GM\dot{M}}{8{\pi}{R}^3{\sigma}}}\Bigl[1-{\Bigl(\frac{R_{*}}{R}}\Bigr)^{1/2}\Bigr]\Bigr\}^{1/4},
\end{equation}
where $M$ is the mass of the WD, $\dot{M}$ is the mass transfer rate,
and $R_*$ is the radius of the WD. $G$ is the gravitation constant and $\sigma$
is the Stefan-Boltzmann constant. This equation defines the standard model for the radial
temperature profile of a CV accretion disk.

The accretion disk is heated by viscous shear. It is now believed that the physical basis
for viscosity is the magnetic instability mechanism \citep{bh91}.
Lack of knowledge of a physical basis for viscosity in the original study of accretion disks
led to introduction of the dimensionless parameter $\alpha$ \citep{ss73}. 
An alternate viscosity prescription is
to use a Reynolds number, typically indicated by the symbol $Re$. 
\citet{p1981} discusses both prescriptions and shows that, for CV
accretion disks that undergo outbursts to have temporal characteristics roughly equal to
observed values, then the $\alpha$ prescription requires values in the range 0.1--1.0, and the 
Reynolds number prescription should use values in the approximate range $10^2$--$10^3$.
Although important advances have been made in specifying the viscosity profile within an
accretion disk annulus (\citet{sb1996,s1996,bh1997,b2002}), it still is not possible to calculate a value of
$\alpha$ or $Re$ from fundamental physical principles to use in modeling a particular annulus. 
An important point is that equation (1) has no explicit dependence on the viscosity 
parameter. We assume that a single value of the viscosity parameter applies to an
entire accretion disk.

We model a CV accretion disk as a nested series of cylindrical annuli.
The theoretical model by \citet[hereafter H90]{hub90} provides the model 
for the individual annuli, with additional development in \citet{hh98} and an earlier
investigation in \citet{kh86}. 
Computer implementation of the model is in TLUSTY (v.200) \citep{h88,hl95}.
The calculation of annulus synthetic spectra from the TLUSTY models uses program
SYNSPEC (v.48) \citep{hsh85,hub90}.  
An annulus synthetic line spectrum 
includes physical line-broadening effects, such as natural,
van der Waals, Stark, or from turbulence. We adopted solar composition for all synthetic spectra.
A useful option in SYNSPEC is to calculate continuum spectra, but also including H and He lines,
rather than line spectra.

H90 develops the theory of the vertical
structure of an annulus. The vertically-averaged viscosity and its specific value as a function
of reference level within the annulus appear explicitly in this theory.	The annulus $T_{\rm eff}$
as a function of radial distance from the geometric center is given by equation (1).

Calculation of a model for a single annulus proceeds in four steps:
(1) calculation of a gray model using mean opacities rather than frequency-dependent ones;
(2) iteration, starting with the gray model but now using frequency-dependent opacities,
while imposing local thermodynamic equilibrium (LTE) excitation relations, 
until changes in a state vector, at all levels, become less than a prescribed small value,
producing a converged model in LTE; (3) relaxing the LTE requirement
and iterating to produce a non-LTE model for the continuum; (4) using the model from (3) and
including line opacities and again iterating to produce a final non-LTE model. 

Our experience
is that models for inner annuli typically converge but convergence becomes more and more fragile
for larger radius annuli until only a gray model can be calculated. Depending on the mass transfer
rate, even the gray model calculations may fail for the largest radius annuli. Based on extensive tests,
we find that plots of annulus synthetic spectra progress smoothly from smaller to larger annulus radii, 
even though
the models transition from LTE to gray models. Consequently, to minimize the required amount of
computer time, we have used gray models exclusively in this investigation.

\subsection{Calculation of synthetic spectra for a CV system}

The BINSYN software suite \citep{linnell96} is used to calculate synthetic spectra for comparison
with the observational data. For an accretion disk system, 
BINSYN produces  separate output synthetic spectra for the total system, the mass gainer, the mass
loser, the accretion disk face, and the accretion disk rim (including a hot spot, if specified).
Line-broadening and line displacement due to 
the Doppler effect
are calculated in BINSYN as the programs generate synthetic spectra for the total system and the
individual system components.
The calculated synthetic spectra require prior assembly of arrays of source synthetic spectra
produced externally to BINSYN, one array for each object (mass gainer, accretion disk face, etc.).
Thus, for a model corresponding to Table~2 (see below) and with a 3500K secondary star, BINSYN would require
21 synthetic spectra corresponding to the annulus radii in Table~2, a synthetic spectrum for the
WD, a synthetic spectrum (in this case duplicating the outer annulus spectrum) for the accretion
disk rim, and a synthetic spectrum for the secondary star. For the latter object, since its
contribution in the present case is negligible, a single synthetic spectrum meets the formal requirement. 
Otherwise, because the secondary star fills its Roche lobe, a range of synthetic spectra would
be required covering the photospheric variation in $T_{\rm eff}$ and log~$g$. The BINSYN model for the
distorted secondary star includes effects of gravity darkening and irradiation by the primary and the
accretion disk. Shadowing of the secondary, by the accretion disk, from irradiation by the primary is
included.

The number
of annuli specified, e.g., in Table~2, must be large enough that interpolation to a non-tabular $T_{\rm eff}$
value can be accurate.
The output spectrum is produced by
interpolation within the array of synthetic spectra to produce, in effect, a local 
photospheric model atmosphere 
at each photospheric segment. 
Integration
over the segments, with proper allowance for Doppler shifts and sources of line broadening,
and with suppression of contributions from segments hidden from the observer, then produces
the synthetic spectrum for the object (star, accretion disk, rim) at the particular orbital 
inclination and longitude under
consideration.
The system synthetic spectrum is the sum of contributions
from the separate objects. 
Stellar objects are represented by the Roche model, including allowance for rotational distortion 
up to critical rotation.

The accretion disk is represented within BINSYN by a specified number (typically 32) 
of concentric annuli with
fixed radial width but of increasing 
thickness (see \citet{linnell96}), up to a specified rim height. Each annulus is divided into 
azimuthal segments (typically 90).
Note that the number of annuli specified
within BINSYN usually exceeds the number of calculated annulus models. This feature is necessary
to provide adequate resolution in calculating, e.g., eclipse effects.
BINSYN calculates a $T_{\rm eff}$ value for each internally specified annulus (the 32 mentioned above);
these are standard model \citep{fkr92} values by 
default, but an option permits assigning an individual $T_{\rm eff}$ to each annulus.
This feature allows evaluation of an arbitrary temperature profile for the accretion disk,
including an isothermal model.
The program optionally permits calculation of irradiative heating of the accretion disk by the WD,
based on a bolometric albedo formalism.	The output spectra from BINSYN are in the same mode as
the annulus, etc., spectra--either line spectra or continuum spectra (i.e., if the synthetic spectra
for the annuli are continuum spectra, that mode having been chosen for the SYNSPEC calculations, then
the system, etc., synthetic spectra produced by BINSTN also will be continuum spectra).

BINSYN requires input specification of the inner and outer radii of the accretion disk.
This feature provides important flexibility to truncate the disk at an inner radius and
to set the outer radius at the tidal cutoff radius as dictated by the mass ratio of the
stellar components. The accretion disk temperature profile typically will
place all of the accretion disk on the "hot branch" required for dwarf novae outbursts
by the Disk Instability Model,
hereafter DIM \citep{osa96}.

\section{WD and secondary star masses and $T_{\rm eff}$ values}

The observed spectra of SDSSJ0809 give no immediate indications of an underlying WD spectrum, 
apparently
preventing a direct determination of the WD $T_{\rm eff}$. 
The subsequent discussion shows that SDSSJ0809 has an accretion disk, so it is
clear that the secondary star must fill its Roche lobe. Recent years have seen major
progress in models of low-mass stars \citep{bar95,bar97,bar98} and recent papers 
\citep{k2000,a2004}
detail the application of these models to the secondary stars in CVs.
From Figures~2 and 3 of \citet{k2000} we determine a secondary component spectral type of
about M3 and a secondary component mass 
${M}_2/{M_{\odot}}=0.30\pm0.10$. This determination has used the theoretical ZAMS, and there
is a substantial difference between observationally determined spectral types and the
theoretical ZAMS in the period range $2-5^{\rm h}$. This difference leads to the large
uncertainty in the secondary component mass. Subsequent sections consider the propagation
of this uncertainty into the system geometry.

\citet{sd98} specify an average WD mass of 0.69$M_{\odot}$
for WDs below the period gap and 0.80$M_{\odot}$ for WDs above the period gap.
Their tabulation includes only one object in the period interval $2^{\rm h}$ to $4^{\rm h}$, and
it is a polar. The systems with periods between $4^{\rm h}$ and $5^{\rm h}$ have WD masses between 0.6$M_{\odot}$
and 1.26$M_{\odot}$, with unreliable values in several cases.
The 192 min period places
the system above the period gap. 
For the purposes of an initial illustration we
adopt a WD mass of $1.0M_{\odot}$, and with an 
uncertainty of $0.2M_{\odot}$.

The lack of deep eclipses sets an upper limit on the orbital inclination. The presence of 
absorption cores associated with emission features (Figure~3),\citep{sz03}, with the absorption cores going
below the continuum level, sets a rough lower limit on the inclination.	We have determined
the orbital inclination for which the secondary component barely fails to eclipse the accretion
disk rim, both for the $q=0.30$ case and for a $q=0.20$ case. For the $q=0.30$ case, the
inclination is $i=65{\degr}$, and for the $q=0.20$ case the inclination is $i=67{\degr}$.
We take the observed low-amplitude light curve, together with the absorption cores in spectral
lines, as evidence of a grazing eclipse of the
accretion disk by the secondary star and adopt an orbital inclination of $i=65{\degr}$.   

Figure~5 shows theoretical radial velocity curves for both cases. The solid curves are the
radial velocity curves for the $q=0.30$ case; the smaller amplitude curve represents the WD
while the larger amplitude curve is the secondary star.
The dashed curve represents the WD for the $q=0.20$ case, with $i=67{\degr}$.
Gamma velocities of -105 ${\rm km}~{\rm sec}^{-1}$ and 
8 ${\rm km}~{\rm sec}^{-1}$ were subtracted from H$\alpha$ 
and H$\beta$, respectively.	These velocities were determined empirically by minimizing the
sum of the residuals from the theoretical radial velocity curve. The curves show the insensitivity
of the system to the value of $q$.
We stress that this paper does {\it not} argue that the H$\alpha$ and H$\beta$
radial velocities track the WD motion in SW Sex systems in general and in this system in particular.

Figure~6 shows a projection view of the $q=0.30$ system. A plot for the $q=0.20$ case is
closely similar to Figure~6.

 \citet{p2000} determine a radius of $r_{\rm wd}=0.00771R_{\odot}$ 
for a WD mass of $1.00{M}_{\odot}$ and a homogeneous 
Hamada-Salpeter carbon model.  For a secondary star mass of 
$0.30{M}_{\odot}$, and assuming the secondary is similar to 
a main sequence star of the same mass,  
we adopt a polar $T_{\rm eff}$ of 3500K for the secondary star \citep{bar98}.
The following analysis will show that the secondary star makes a negligible contribution
to the system shortward of 7000\AA.
   
Several SW Sex stars have high WD $T_{\rm eff}$ values. Examples include DW UMa, 
40,000K--50,000K \citep{kn2000,ab2003,ho2003},
and UUAqr, 34,000K \citep{bsc1994}. Similarly, MV Lyr, a well-known novalike CV, has a $T_{\rm eff}$ of 
47,000K \citep{ho2004}.
We chose a test $T_{\rm eff}$ of 35,000K for SDSSJ0809, with the possibility in mind of a subsequent
revision upward (or downward).

 \section{Initial estimate of mass transfer rate and the accretion disk tidal cutoff radius 
 }

The spectroscopic similarity of SDSSJ0809 to an SW Sex 
system also suggests our initial choice of mass transfer rate.  
SW Sex stars (for reviews see \citet{h2000,g2000,ho2003}) are 
novalike (NL) objects with mass transfer rates above 
the critical rate to avoid disk outbursts \citep{osa96}.  
From \citet{osa96}, the critical mass transfer rate is 
$\dot{M}_{\rm crit}{\cong}4.28{\times}10^{-9}(P_{\rm orb}/4^{\rm hr})$ 
$M_{\odot} {\rm yr^{-1}}$.  In the case of SDSSJ0809, the Osaki 
expression gives a critical rate of 
$3.4{\times}10^{-9}M_{\odot}/{\rm yr}^{-1}$, but we expect 
there may be appreciable case-to-case variation.  
\citet{pat84} predicts a mass transfer rate of 
$2.5{\times}10^{-10}M_{\odot} {\rm yr^{-1}}$ at the 
top of the period gap, while \citet{ham88} determine a value 
of $1.0{\times}10^{-9}M_{\odot} {\rm yr^{-1}}$. We discuss 
annulus models for four mass transfer rates in the 
following sections.

A number of studies have considered the tidal cutoff 
boundary, $r_{\rm d}$, of accretion disks 
(\citet{pac77,pp77,w88,scr88,wk91,g93,warner95}). 
Equation 2.61 in \citet{warner95}, gives a tidal cutoff radius of 
$0.49D$, 
where $D$ is the separation of the stellar components. The other 
authors cited agree on 
$r_{\rm d}{\cong}0.33D$.  
The latter prescription gives a tidal cutoff radius of 
$51r_{\rm wd}$, 
where $r_{\rm wd}$ is the WD radius. In this connection, the difference
in $D$ between a $q=0.30$ system and a $q=0.20$ system is relatively
small. Consequently, the cutoff radius of the accretion disk is
insensitive to $q$.

CVs whose mass transfer rates are above the critical rate, and so are stable, 
are expected to have an accretion disk temperature profile 
given by the standard model \citep{fkr92}.  \citet{sm82} 
finds that the critical temperature for instability in the 
DIM model \citep{las2001} 
occurs at $T_{\rm eff}{\cong}6300$K.  
The standard model with the 
$r_d=0.49D$ 
cutoff radius produces outer annuli $T_{\rm eff}$ values 
which are below the critical temperature for instability.  The 
$r_d=0.33D$ 
cutoff radius produces an outermost annulus $T_{\rm eff}$ value 
that is at or above the critical temperature for a mass transfer rate
of at least $1.0{\times}10^{-9}M_{\odot} {\rm yr^{-1}}$. This result agrees with 
the study by \citet{sm83}, that 
NL systems 
have outer radii whose $T_{\rm eff}$ values lie above the 
critical temperature.  (We thank J.-P. Lasota for a 
communication on this point.) We adopt the tidal cutoff radius 
$r_d=0.33D$ 
for this system, and note that the WD mass and system mass 
ratio are uncertain, with a corresponding uncertainty in the 
exact cutoff radius.

\section{Annulus models for four mass transfer rates}

Without independent information on the mass transfer rate, we used rates typical for
systems with periods between 3-4 hrs.
To this end we have calculated standard models
for a range of mass transfer rates characteristic of SW Sex stars and have found the best-fitting model
for each mass transfer rate. 
We calculated four sets of accretion disk annuli with the
adopted WD mass of $M=1.0M_\odot$. 
We chose
mass transfer rates of $\dot{M}$=$1.0{\times}10^{-9}M_{\odot}{\rm yr}^{-1}$, 
$2.0{\times}10^{-9}M_{\odot}{\rm yr}^{-1}$, 
$3.0{\times}10^{-9}M_{\odot}{\rm yr}^{-1}$, and $5.0{\times}10^{-9}M_{\odot}{\rm yr}^{-1}$.	
Table 2 lists annulus properties for the mass
transfer rate we ultimately chose as most likely.

The TLUSTY default value for viscosity is $\alpha=0.1$, and since this is the value widely used for the
``hot branch" in the DIM, we adopt it \citep{p1981,s2000,t2000,las2001}. 
The viscosity prescription strongly affects
the annulus vertical structure, including the annulus optical thickness 
\citep{kh86,hub90}.
The TLUSTY parameters ${\zeta}_0$ and ${\zeta}_1$ control the vertical
viscosity profile in an annulus. 
We have used the default values which assure that there
is no``thermal catastrophe" \citep{hub90,hh98}. The ``thermal catastrophe" results when
the vertical viscosity profile requires appreciable energy dissipation in low density
layers where cooling from strong resonance lines of light metals becomes important.

The TLUSTY models all produce a standard model $T_{\rm eff}$ \citep{fkr92} for an individual 
annulus. The standard model requires all of the thermal energy generated
within an annulus to result from viscous dissipation. 
Our current models also include irradiation of the disk by the WD, which 
can modify the
standard model $T_{\rm eff}({\rm R})$ relation. 

The chemical composition of the inner annuli, extending to a transition radius described below, include
H and He as explicit atoms, 
and the remaining first 30 atomic species as implicit atoms. 
Inclusion as an implicit atom means that the atomic
species contributes to the total number of particles and the total charge, but not to
the opacity.
The vertical temperature profile of a given small radius annulus varies
smoothly from the central plane to the upper boundary of the annulus. At larger radii the temperature
profiles show increasingly steep drops toward the upper boundary. For these annuli we 
``switched on" convection in TLUSTY. 
In addition, the larger radius annuli
have lower $T_{\rm eff}$ values, and opacity due to metals is included.
Beginning at $T_{\rm eff}{\thickapprox}12000$K,
our annulus synthetic spectra include H, He, C, Mg, Al, Si, and Fe as explicit atoms
and the remaining first 30 atomic species as implicit atoms. 
Because they have fairly high-lying first excited states, N and O do not contribute
importantly to the continuum opacity of interest (photoionization from the ground state
is close to the Lyman limit, indeed shortward of it for O). 
 
The data for a given annulus have been extracted from an extensive TLUSTY tabulation of
physical properties at all (typically 70) reference levels in the annulus, from the
central plane to the upper boundary. Note from the last column of Table~2 that all 
annuli are optically
thick. This condition, that all annuli are optically thick, is also true for all 
of the other mass transfer rates calculated. 

The outer annuli for the
mass transfer rate of  $1.0{\times}10^{-9}M_{\odot} {\rm yr^{-1}}$
have outermost
annulus $T_{\rm eff}$ values that fall below the stability limit for the DIM model. 
Smaller mass transfer rates	have appreciably larger fractions of the outer annuli that fall
below the stability limit.
This sets a lower
limit on the mass transfer rates to consider.

\mathversion{bold}
\section{System models for successive mass transfer rates}
\mathversion{normal}

We began with a standard model $T(R)$  for a mass transfer rate of $1.0{\times}10^{-9}M_{\odot}~{\rm yr^{-1}}$
and an untruncated accretion disk.
Because the $FUSE2$ exposure was closer to the time of the $HST$ exposure than the $FUSE1$ exposure, 
we initially chose
the $FUSE2$ spectrum for comparison with our simulations.
(The number of plots required to show the results for all four mass transfer rates would be very large.
Consequently we present illustrative plots for only one of the mass transfer rates. Similarly, we tabulate 
annulus properties for only one of the four mass transfer rates (Table~2)).
A continuum spectrum fit to the $FUSE2$, $HST$ and SDSS spectra for the untruncated disk had far
too much UV flux as compared with the $HST$ and $FUSE2$ spectra. This implied too large a contribution
from high $T_{\rm eff}$ annuli. From equation~(1), the only way to reduce those contributions, and
maintain the standard model, is to truncate the inner accretion disk.
Accordingly, we started with
the outer radius at the tidal cutoff radius and calculated models truncated at inner radii of 
1.0, 4.0, 6.0, 7.5,	8.2, 9.0,
and 10.0 times the WD radius. 
We calculated continuum spectra for the models
and determined individual scaling factors to fit them to the SDSS spectrum, requiring accurate superposition
among the models at the long wavelength end of the SDSS spectrum.
A model truncated at 8.2 times the WD radius was the best overall fit to the $FUSE2$, $HST$, and SDSS spectra,
as judged visually. It would be very difficult to be quantitative in this estimate, since the degree of
departure from a good fit in the optical must be weighed against what is an approximate fit to the $HST$ and
$FUSE$ spectra. In addition, the many UV emission line features make location of the continuum problematic.
The illustrative plots explain this difficulty.
The synthetic spectrum spectral gradient was clearly too shallow in the fit to the SDSS spectrum, and was slightly
too hot in the UV. The UV discrepancy increased on substituting the $FUSE1$ spectrum. Since the $FUSE1$
spectrum provides a more stringent UV constraint on the models (setting a maximum WD contribution), 
and since the $FUSE1$ spectrum mates fairly smoothly
with the $HST$ spectrum in their overlap region, while the $FUSE2$ spectrum does not, we decided to use
the $FUSE1$ spectrum in our final comparisons between the models and the observed spectra. It is
possible that the higher excitation $FUSE2$ spectrum results from a higher mass transfer rate from
the secondary star.

A slight increase in the truncation radius, to 9.0 times the WD radius, made the fit to the optical spectrum
slightly worse without a great improvement in the UV fit. Since the system synthetic spectrum had too much UV
flux, and since increasing the truncation radius did not help, we tested the other option of reducing 
the WD $T_{\rm eff}$.
Substituting a 20,000K WD reduced the WD
contribution to the synthetic spectrum to a nearly negligible value, and improved the UV fit.

A mass transfer rate of $2.0{\times}10^{-9}M_{\odot}~{\rm yr^{-1}}$	presented the same problem for an
untruncated accretion disk.
Following the procedure of the previous mass transfer rate,
we started with
the outer radius at the tidal cutoff radius and calculated models truncated at inner radii of 
1.0, 4.0, 6.0, 8.0,	9.7,
and 12.0 times the WD radius. 
The 9.7 model produced a very good fit to the optical (SDSS) spectrum but was too hot in the UV. As
with the first mass transfer rate, substituting a 20,000K WD substantially improved the UV fit.

We next considered a mass transfer rate of $3.0{\times}10^{-9}M_{\odot}~{\rm yr^{-1}}$.
We started with
the outer radius at the tidal cutoff radius and calculated models truncated at inner radii of 
1.0, 4.0, 8.0, 12.0, 14.3, 16.0,
and 18.0 times the WD radius.
Figure~7 shows the superposition of these models on the optical spectrum. Note the progression from
a too large spectral gradient to a too small gradient for the synthetic spectra. It clearly is
possible to select a best fitting model in this spectral region, by visual inspection, from among 
the available choices. Although we have a $\chi^2$ fitting procedure available, we do not believe
its use is warranted in the present illustration.
Figure~8 connects the optical region to the UV. Note the large separation of the synthetic spectra 
in the UV for relatively small differences in the optical. Figure~9 shows the fits in the UV only.
The best overall fit appears to be the model with inner truncation radius $r=16.0r_{\rm wd}$.
(For brevity, the subsequent text will use ``the 16.0 model" to reference this model, and similarly
for other inner truncation radii.) 
The 14.3 model is too bright in the 1100\AA~to 1350\AA~interval while the 18.0 model has a large
flux discrepancy longward of 1250\AA. Figure~10
shows the fit of the 16.0 model to the optical spectrum, and Figure~11 shows the fit to the UV spectra.
The Figure~10 result, that the contribution of the secondary star is essentially negligible in the optical,
and completely so in the UV, applies to all of the models in this investigation.
The fit of the 14.3 model to the optical spectrum is nearly identical to Figure~10. From
Figure~7, the 18.0 model (the lowest of the seven plots) has a clearly too small spectral gradient
in the optical. This discussion illustrates how we can use visual estimates to select the best fitting
model. Plots for other mass transfer rates show characteristics comparable to those shown here but with
specific differences described for those mass transfer rates.

For comparison, Figure~12 substitutes the $FUSE2$ spectrum.
Note that, in Figure~11, the $FUSE1$ spectrum and the $HST$ spectrum fit reasonably smoothly in the overlap
region, 1150\AA~to 1200\AA, while there is a clear discontinuity in the Figure~12 overlap. This
comparison justifies our use of the $FUSE1$ spectrum. An uncertainty remains for the SDSS spectrum
because of its non-simultaneity with the $HST$ or $FUSE$ spectra.

The fit to the $FUSE1$ spectrum shown in Figure~11 places the synthetic spectrum too high in the vicinity
of 1150\AA. Figure~13 shows the result of substituting a 20,000K WD. The WD contribution, barely
visible at the bottom, now is nearly negligible. The fit near 1150\AA~is improved but the flux
deficit longward of 1300\AA~is larger. 

The same procedure was followed for the mass transfer rate of $5.0{\times}10^{-9}M_{\odot}~{\rm yr^{-1}}$. 
We calculated a set of five models which bracketed the $HST$ spectrum from too much to too little flux,
with the synthetic spectra normalized to agree with the optical spectrum at 9200\AA. 
All of these models showed a too large spectral gradient on comparison
with the optical spectrum. 
All of the models were too hot
in the $FUSE$ range, for a 35,000K WD. Substituting a 20,000K WD improved the UV fit but in the $FUSE$
range the synthetic spectrum still was too hot. With this substitution, as with
the other mass transfer rates, the WD contribution became negligible.  

Larger mass transfer rates would produce still greater discrepancies in the fits.
Consequently, we do not consider larger mass
transfer rates.	The best-fitting model at the minimum mass transfer rate, 
$1.0{\times}10^{-9}M_{\odot}~{\rm yr^{-1}}$, is clearly inferior to the best-fitting models
at $2.0{\times}10^{-9}M_{\odot}~{\rm yr^{-1}}$ and $3.0{\times}10^{-9}M_{\odot}~{\rm yr^{-1}}$,
which are nearly comparable. The best-fitting model at the maximum mass transfer rate,
$5.0{\times}10^{-9}M_{\odot}~{\rm yr^{-1}}$, also is inferior to the two intermediate
mass transfer rates. We judge that the $3.0{\times}10^{-9}M_{\odot}~{\rm yr^{-1}}$
mass transfer rate is a slightly better spectral fit than the
$2.0{\times}10^{-9}M_{\odot}~{\rm yr^{-1}}$ rate and adopt the mass transfer rate of
$3.0{\times}10^{-9}M_{\odot}~{\rm yr^{-1}}$ for this system.

The most important result from the analysis of
the four mass transfer rates is that the accretion disk must be truncated if the standard model
applies. 
The fits to the $HST$ spectrum
exhibit a flux deficiency between about 1400\AA~and the 1700\AA~limit of the $HST$ spectrum for
models that have acceptable flux levels in the $FUSE$ range. We see no way to accommodate that
discrepancy by maintaining the standard model. 
If this result is to be accepted, then a physical effect must be introduced to produce the
truncation. The two possibilities are evaporation (\citet{me2000,dubus2001,las2001}) and interaction 
with a WD magnetic field. 
Using the evaporation equation of \citet[eq.3]{me2000}, and setting the evaporation rate equal to the mass
transfer rate,
it can be shown that evaporation is too small an effect by four orders of magnitude. 
\citet{g2000} state
that the SW Sex phenomenon can be explained in the context of a non-magnetic WD, and \citet{r2001} identify
circular polarization in the SW Sex star LS Peg, and they argue that magnetic accretion plays a fundamental 
role in SW Sex stars. 
\citet{ho2003} summarize and describe the various arguments and evidence for and against magnetic 
WDs in SW Sex stars. Because of this uncertainty, and in anticipation of the results presented
in \S12, we do not discuss possible truncation of the accretion disk by a magnetic field.

A further important result from the 
truncation models is that the best fit to the $FUSE$ spectrum requires a nearly negligible WD
contribution to the system synthetic spectrum.
This result is hard to understand, since
Figure~6 clearly shows that the inclination is too small for a vertically enlarged accretion disk
rim to hide the WD. Hiding the WD would be even more difficult if the accretion disk is truncated.
The negligible WD contribution cannot be a result of the accretion disk hiding the WD.
The truncation models cover the full range of acceptable mass transfer rates, and a cool WD
produces a better synthetic system spectrum fit than the 35,000K WD in all cases.
The alternative to hiding the WD behind the accretion disk rim, that the WD actually is cool, 
is difficult to understand in view of the large
mass transfer rate which would be expected to produce a hot WD \citep{sion99,szkody02,ab2003}.
The spreading layer theory
applied to WDs \citep{pb04} might permit a prediction of the WD $T_{\rm eff}$ for given $M_{\rm WD}$
and $\dot{M}$.	
Although we have found a model
which is a fairly good fit to the combined spectroscopic data, the model appears unrealistic and
we reject the truncated accretion disk scenario.

We are forced to
conclude that the accretion disk in SDSSJ0809 may depart from a standard model, which is the
basis of all the models we have calculated to this point.

\section{Non-standard model accretion disks}

Various studies (e.g., \citet{ru1992}), based on image reconstruction, find a flatter temperature 
profile than the standard 
$T(R)$ relation. It is of interest to adopt an isothermal
accretion disk and test whether that prescription produces a viable model. 
Figure~14 shows the fits of isothermal 15,500K, 13,000K, and
12,000K untruncated accretion disks to the observed optical spectrum. 
Figure~15 shows the fits to the $HST$ and $FUSE1$ spectra. 
Each model includes the contribution of
a 35,000K WD. The WD contribution, normalized as for the 13000K model, is shown separately at the bottom
of Figure~14 and Figure~15. 
The
12,000K model fits the $HST$ spectrum fairly well but has too little flux in the $FUSE$ region.
All three models are 
poor fits to the observed
optical spectrum.

The tomographic image reconstruction for SW Sex itself (\citet{ru1992}) shows that the accretion disk
temperature profile follows the standard model in the outer part of the accretion disk, but at a
transition radius the temperature profile becomes flat and remains nearly so to the	inner edge, 
with a slow
increase.
We have implemented a model similar to this reconstruction by first adopting the 
mass transfer rate, $3.0{\times}10^{-9}M_{\odot}~{\rm yr^{-1}}$, for the best standard
model, and then finding the radius for which the local $T_{\rm eff}$ was 13,000K.
This transition value was based on our result for the isothermal disk models. 
We used an accretion disk that was not 
truncated and set the $T_{\rm eff}$ of all annuli within the transition radius equal to 13,000K.
The resulting model, in contrast to the isothermal models, fits the optical spectrum very well.
However, the fit to the $HST$ spectrum left appreciable positive residuals.
We experimented with a larger temperature isothermal region, with a smaller radius for the
crossover to the standard model. Eventually we found that an isothermal region of 14,000K, with
a crossover to the standard model at $r/r_{\rm wd}=18.0$, gave a fair representation of the
$HST$ and $FUSE1$ spectra. (Note that the tidal cutoff radius is at $r/r_{\rm wd}=51.3$). The
temperature profile of the accretion disk is in Table~5.
This model assumes the accretion disk extends to the WD equator, so only half of the WD is visible.
The BINSYN software allows for eclipse of inner annuli segments that the WD eclipses.

The fit
to the $HST$ and $FUSE1$ spectra suggest the possibility of an improved fit with a higher $T_{\rm eff}$
WD. We calculated models, in addition to the 35,000K model, of 40,000K, 45,000K, and 50,000K. All four
models have nearly identical fits to the upper (discovery) SDSS spectrum, as shown in Figure~16. 
The lower (archival) SDSS
spectrum is discussed in the following section. 
Figure~17 shows
the fits to the $HST$ and $FUSE1$ spectra. The bottom spectrum is the contribution of the accretion disk.
Based on the fit near 900\AA,
the 45,000K model roughly bisects the observed continuum.

The synthetic spectra are given in Eddington flux
units; a divisor, $s_N$, is applied to superpose a synthetic spectrum on the observed spectra.
Since the observed flux is tabulated in units of 
$1.0{\times}10^{-14}~{\rm erg}~{\rm s}^{-1}~{\rm cm}^{-2}~{\rm \AA}^{-1}$,
the distance, $d$, is given by $d^2=s_N/1.0{\times}10^{-14}$. In the present case, for the
model using the discovery SDSS spectrum, the
scaling factor is $6.1{\times}10^{28}$; the formal derived distance to SDSSJ0809 is 800pc.
Note that there is a large associated uncertainty.
Tentative system parameters are listed in Table~3.
The Table~3
parameters are consistent with  other SW Sex systems. 
 
\section{The archival SDSS spectrum and model implications}

Figure~16 shows the fit of the Table~5 model to the discovery SDSS spectrum. The lower SDSS spectrum
is the archival spectrum. Dividing the discovery spectrum by 1.3 and overplotting produces a close
fit to the archival spectrum between 5500\AA~and 9200\AA. The discovery spectrum shows increasing
positive residuals shortward of 5500\AA, reaching a maximum at 4100\AA, and decreasing to
coincidence with the archival spectrum at 3800\AA.

Before attempting to model the archival SDSS spectrum we seek a likely physical basis for the observed
change from the discovery spectrum. An important point is that the change took place in only two days.
In \S11 we have argued that the accretion disk cannot be approximated by a standard model and that
models truncated on the inner edge are unrealistic. In \S12 we found that isothermal accretion disks
produce poorly-fitting models, but that a model whose temperature profile mimics SW Sex satisfactorily
fits the discovery SDSS spectrum as well as the UV data. In \S11 we have already suggested that the
$FUSE2$ spectrum differs from the $FUSE1$ spectrum because of a higher mass transer rate. We believe
the two SDSS spectra most likely differ for the same reason. 

If the mass transfer rate from the secondary star decreases, a cooling wave will propagate inward from
the outer accretion disk rim. The temperature of the outer annuli could fall below the transition
temperature from the high state. With this scenario in mind, we started with the Table~5 profile,
reduced the outer annuli temperatures
and calculated a new system model. Although the flux level was reduced, the system synthetic spectrum
had a spectral gradient that was too high compared with the archival spectrum.
The scaling
factor to superpose the synthetic spectrum on either observed spectrum has to remain fixed, since it
connects directly to the system distance. 
We continued temperature reductions for all annuli, reaching the final temperature profile of
Table~6. 
Note that a
complete system model was required at each new accretion disk temperature profile. 
The final fit to the archival
spectrum is in Figure~18. Note that Figure~18
presents fits for the same four values of the WD temperature as in Figure~16.
Figure~19 shows the four model fits to the UV spectra. 
Note the large reduction in the contribution of the accretion disk as shown by
the lowest plot.

These fits have unacceptably large residuals from
the $HST$ spectrum. 
A basic problem is that we have only a single $HST$ spectrum and it is not contemporaneous with any of the
other observed spectra. No information is available about whether the $HST$ spectrum changes in a time interval
of days, and by how much.

The proposed scenario to explain the differences in the two SDSS spectra resembles the widely-accepted 
explanation of VY Sculptoris star 
behavior \citep{livio94,king98}. However, SW~Sex objects do not undergo the very large brightness
variations of VY Scl systems, so the physical cause of mass transfer variation in SDSSJ0809 is
uncertain. A detailed study of the high-state/low-state transition in the VY~Scl system V794~Aql 
\citep{hon94} provides a
useful comparison. The drops in V794~Aql brightness have rates that exceed the two-day drop of 0.3 mag.
for SDSSJ0809, so the SDSSJ0809 drop rate is not excessive. The theoretical time-dependent model for 
V794~Aql 
shows two-day changes in annulus 
$T_{\rm eff}$
values as large as 10,000K. Also, successive two-day log~$T_{\rm eff}$ profiles across the accretion disk
are nearly vertical displacements from each other except near the end of the simulation. Thus, the
changes from Table~5 to Table~6 appear to be theoretically acceptable, based on a comparison 
with V794 Aql. The two cases are not entirely comparable since the SDSSJ0809 starting model (Table~5)
is not a standard model.

Although it originally appeared reasonable to calculate a system model fitting the combined
spectra then available, we conclude
that contemporaneous spectra are essential for reliable models of systems which may show significant
day-to-day changes. Note that this conclusion is based on a quantitative comparison
with model calculations. 

\section{Discussion}

The $FUSE1$ spectrum fits smoothly
with the $HST$ spectrum in the overlap region, while the $FUSE2$ spectrum does not fit
well. This is inconclusive evidence that SDSSJ0809 was in a lower accretion state at the times of
both the $HST$ and $FUSE1$ exposures, and that the $FUSE2$ spectrum was obtained during a higher state.
Because of this uncertainty, we emphasize again that this study has been an illustration of
techniques to simulate accretion disks, with the SDSSJ0809 observational data serving as
the vehicle for that illustration.

All of these TLUSTY annulus models are optically thick, including the nonstandard models.
The optically-thick annulus spectra with their absorption lines and large
Balmer discontinuities are in conflict with all of the observed emission line spectra
({\it FUSE}, STIS, and SDSS).
What do the emission lines tell us and what is the most likely way the model should
be changed to produce the observed emission lines? 
All of the SDSSJ0809 emission lines are single-peaked, one of the defining characteristics
of SW Sex stars. 
In contrast to the SW Sex result
that the emission lines arise exclusively from the accretion stream shock \citep{g2000},
\citet{h2000} argues for emission lines produced in a wind. Hellier asserts that a wind
provides the most straightforward explanation of single-peaked emission lines, and that 
in particular the wind component
fills in any double-peaked disk emission in the Balmer lines.
(Also see \citet{dp2000}).
Table~4 provides some details
concerning the Balmer emission lines in SDSSJ0809. 
There are no P Cygni profiles
among the SDSSJ0809 Balmer lines or He lines. There is only a hint of an absorption 
component at the blue edge of the C IV 1550\AA~lines. 

The most direct explanation of the
SDSSJ0809 emission lines is production in a chromospheric or coronal region overlying
the accretion disk.
Several mechanisms have been
proposed that could produce coronae: \citet{mm94} use a coronal siphon; \citet{ml92} depend
on sound waves that accelerate to form shocks; \citet{lp77} propose general nonthermal
processes to transport energy vertically from the accretion disk;
\citet{hub90} depends on the vertical
variation of viscosity within an annulus; \citet{c2000} applies a thermal evaporative
instability \citep{s1986}. 

An extensive literature exists relating to formation of emission lines in CV systems.
\citet{w1980,t1981} and \citet{cl89} discuss emission line formation in optically thin outer
annuli of accretion disks. 
\citet{ko1996} discuss emission line formation as a result of X-ray/EUV illumination of the
accretion disk by the central object. 
\citet{hm86} and \citet{f1997} discuss emission line formation in more
general cases. 
The SDSS spectrum does not extend far enough in the UV to
determine the last resolvable Balmer line; hence, it is not possible to apply the
Inglis-Teller equation to determine the electron density where the continuum is
produced. 
Note from Table~4 that the Balmer decrement is relatively flat.
(Compare with case A and case B for planetary nebulae, Table~2 of \citet{al1968}, and
the result that Balmer decrements for some Be stars approximate those of planetary
nebulae (\citet{bb1953} and references therein).)
\citet{hm86} point out that in the case of a flat Balmer decrement the Balmer emission 
lines are optically thick.
The fact that all of the annulus models are optically thick, together 
with the result
that the Balmer emission lines are optically thick, is consistent with an origin of emission
lines in a chromosphere or corona, not in optically thin outer annuli. 
Moreover, emission lines produced in optically thin outer regions of an accretion disk
occur in a temperature environment of 6000K--9000K \citep{cl89}. The high excitation
emission lines in SDSSJ0809, particularly in the {\it FUSE} spectra, require a much higher
temperature. 
A notable feature of the observed optical Balmer emission lines is that they are narrower than the
synthetic spectrum absorption lines (e.g., see Figure~10). The model absorption lines are
broadened by Keplerian rotation appropriate to the orbital plane. The
narrower emission lines could be an indication they are produced well above the
accretion disk face, where the horizontal acceleration due to the WD, and the associated
Keplerian speed, is smaller.

Extrapolation of the SDSS spectrum indicates the absence of a Balmer jump.
\citet{lad1989} points out that this condition implies a much lower vertical temperature 
gradient in an accretion disk than is true in a stellar atmosphere, and likely associates
with energy production in the accretion disk, in contrast to a stellar atmosphere.

We doubt that addition of an emission
line slab to any of the models will smoothly fill in the absorption lines and convert the
absorption line spectrum to the observed emission line spectrum. Rather, we believe that
a modification of the individual annulus models is necessary to convert their
spectra to emission line spectra. What are the prospects to do this?
\citet{hub90} provides an analytical model for accretion disk annuli that is the
basis of the TLUSTY annulus model calculation. As discussed in \S10, TLUSTY includes
two parameters, ${\zeta}_0$ and ${\zeta}_1$, that specify the vertical viscosity
profile in an annulus, and therefore the rate of energy generation as a function of
$z$ value. These parameters have default values that prevent a ``thermal
catastrophe" \citep{hub90,hh98}. Carefully modified values of these parameters can, in principle, produce
emission lines. 

The \citet{hub90} analytical model specifically considers conditions that lead to
boundary temperatures higher than the annulus $T_{\rm eff}$.
As discussed by \citet{hub90}, the energy dissipated in low density (coronal) layers 
can be removed
by strong (emission) resonance lines of abundant species, such as H I, Mg II, C II, Al III, Si IV,
C IV, N V, O VI, etc. Numerical simulations for actual physical models \citep{s1996} 
indicate that viscosity
increases toward the surface, as required for the proposed TLUSTY models. 
\citet{t2000} discusses our present understanding of viscosity, while \citet{h2001}
discusses the present status of magnetohydrodynamic simulations of cylindrical
Keplerian disks and their relation to viscosity.
The calculation of TLUSTY models with modified values of ${\zeta}_0$ and ${\zeta}_1$ is 
beyond the scope of
the present investigation.

We have not accounted for the half of the liberated potential energy associated with the mass
transfer stream that is usually assigned to accretion luminosity. Part could be dissipated in the
shock where the transfer stream merges with the accretion disk. Although BINSYN has the
capability to model a rim bright spot, the light curve provides no indication of a bright
spot's existence.
A major part of the energy goes into heating the WD.
We assume the remainder is carried away in a wind.

An original motivation for analyzing SDSSJ0809 was the hope that visibility of the WD could
be demonstrated.
The primary argument for detection of the WD is the model's approximate representation of the 
UV continuum shortward of 1200\AA.
If more extensive observational data support this model, then SDSSJ0809 is 
very likely an SW Sex star with a high WD $T_{\rm eff}$
in agreement with $T_{\rm eff}$ values for other SW Sex stars.

\acknowledgements

The authors thank the anonymous referee for detailed comments; addressing them has greatly 
improved the presentation.
The research described in this paper was carried out, in part, at 
the Jet Propulsion Laboratory, California Institute of Technology, 
and was sponsored by the National Aeronautics and Space Administration.  
P.S., D.W.H., K.S.L., and A.P.L. are grateful for support from NASA {\it FUSE} grant NAG5-13656, 
NASA {\it HST} grants GO-09357.06, GO-09724, and AR-10674, and NSF grant AST 02-05875.
BTG was supported by a PPARC Advanced Fellowship.



\begin{deluxetable}{ccccc}
\tablenum{1}
\tablewidth{0pt}
\tablecaption{{\em FUSE} Observation Log \label{t-fuselog}}
\tablehead{
\colhead{Observation} &
\colhead{Exposure} &
\multicolumn{2}{c}{Start Time} &
\colhead{Total Exposure} \\
\colhead{\#} &
\colhead{\#} &
\colhead{(UT)} &
\colhead{(HJD$-$2450000)} &
\colhead{(s)} 
}
\startdata
 1 & 1 & 28 Mar 2003 11:36:10 & 2726.985218 & 1590 \\
   & 2 & 28 Mar 2003 13:01:23 & 2727.044391 & 2473 \\
   & 3 & 28 Mar 2003 14:40:37 & 2727.113297 & 2515 \\
   & 4 & 28 Mar 2003 16:29:19 & 2727.188776 & 1888 \\ \\

 2 & 1 & 16 Mar 2004 14:44:13 & 3081.116762 & 3257 \\
   & 2 & 16 Mar 2004 16:44:41 & 3081.200413 & 2023 \\
   & 3 & 16 Mar 2004 18:31:30 & 3081.274585 & 1606 \\
   & 4 & 16 Mar 2004 20:09:14 & 3081.342450 & 1737 \\
   & 5 & 16 Mar 2004 21:53:07 & 3081.414585 & 1495 \\
   & 6 & 16 Mar 2004 23:02:56 & 3081.463065 & 2050 \\
   & 7 & 16 Mar 2004 23:36:45 & 3081.486547 & 1272 \\
   & 8 & 17 Mar 2004 00:43:58 & 3081.533221 & \phn560 
\enddata
\tablecomments{	``Total Exposure" will be longer than the
final usable exposure time because of data that are rejected
for quality reasons during the standard pipeline processing.
} 
\end{deluxetable}


\begin{deluxetable}{rrrrrrrr}
\tablewidth{0pt}
\tablenum{2}
\tablecaption{Properties of accretion disk with mass transfer rate 
$\dot{M}=3.0E(-9){M}_{\odot}{\rm yr}^{-1}$ and WD mass of $1.0{M}_{\odot}$.}
\tablehead{	  
\colhead{$r/r_{\rm wd}$} & \colhead{$T_{\rm eff}$} & \colhead{$m_0$} 
& \colhead{$T_0$} & \colhead{log~$g$}
& \colhead{$z_0$} & \colhead{$Ne$} & \colhead{{$\tau_{\rm Ross}$}}}
\startdata
1.36  &  66333  &  8.967E3   &  54213  &  7.16   & 4.63E7  & 1.11E15  & 1.48E4\\
2.00  &	 59475  &  1.095E4	 &  48448  &  6.90   & 8.01E7  & 6.75E14  & 1.76E4\\
3.00  &	 48093  &  1.070E4	 &  39159  &  6.60   & 1.35E8  & 4.14E14  & 2.08E4\\
4.00  &	 40423  &  9.959E3	 &  32916  &  6.38   & 1.94E8  & 2.90E14  & 2.39E4\\
5.00  &	 35062  &  9.251E3	 &  28556  &  6.21   & 2.55E8  & 2.24E14  & 2.71E4\\
6.00  &	 31106  &  8.639E3	 &  25342  &  6.07   & 3.18E8  & 1.82E14  & 3.06E4\\
7.00  &	 28058  &  8.115E3	 &  22867  &  5.95   & 3.81E8  & 1.52E14  & 3.39E4\\
8.00  &	 25630  &  7.665E3	 &  20894  &  5.84   & 4.47E8  & 1.31E14  & 3.77E4\\
10.00 &	 21986  &  6.934E3	 &  17395  &  5.67   & 5.84E8  & 1.02E14  & 4.58E4\\
12.00 &	 19367  &  6.365E3	 &  15812  &  5.53   & 7.24E8  & 8.21E13  & 5.42E4\\
14.00 &	 17381  &  5.907E3	 &  14201  &  5.40   & 8.68E8  & 6.74E13  & 6.25E4\\
16.00 &	 15816  &  5.529E3	 &  12933  &  5.30   & 1.01E9  & 5.71E13  & 7.00E4\\
18.00 &	 14547  &  5.210E3	 &  11905  &  5.20   & 1.16E9  & 4.96E13  & 7.60E4\\
20.00 &	 13495  &  4.937E3	 &  11054  &  5.12   & 1.31E9  & 4.36E13  & 7.98E4\\
22.00 &	 12606  &  4.700E3	 &  10333  &  5.04   & 1.44E9  & 3.78E13  & 8.13E4\\
24.00 &	 11843  &  4.492E3	 &   9717  &  4.97   & 1.59E9  & 3.25E13  & 8.10E4\\
30.00 &	 10086  &  3.993E3	 &   8287  &  4.78   & 2.02E9  & 1.49E13  & 7.52E4\\
34.00 &	  9213  &  3.734E3	 &   7572  &  4.68   & 2.32E9  & 6.68E12  & 7.17E4\\
40.00 &	  8188  &  3.420E3	 &   6744  &  4.61   & 3.13E9  & 1.87E12  & 6.97E4\\
45.00 &	  7516  &  3.207E3	 &   6202  &  4.39   & 2.68E9  & 5.27E11  & 6.69E4\\
50.00 &   6961  &  3.026E3   &   5759  &  4.18   & 2.36E9  & 1.59E11  & 7.48E4\\
\enddata
\tablecomments{Each line in the table represents a separate annulus.
The column headed by $m_0$ is the column mass above the central plane.
The column headed by $T_0$ is the
boundary temperature. Compare with $T_{\rm eff}$. 
The log~$g$ values are at an optical depth of 0.9. 
The $z_0$ column gives the height of the
annulus in cm. 
The accretion disk radius at the tidal cutoff boundary is 
$2.75{\times}10^{10}{\rm cm}$.
The $Ne$ column is the electron density at the upper annulus boundary.
The $\tau_{\rm Ross}$ column is the Rosseland optical depth at the central plane.
A viscosity parameter $\alpha=0.1$ was used for all annuli.}		 
\end{deluxetable}

\begin{deluxetable}{llll}
\tablewidth{0pt}
\tablenum{3}
\tablecaption{SDSSJ0809 Model System Parameters}
\tablehead{
\colhead{parameter} & \colhead{value} & \colhead{parameter} & \colhead{value}}
\startdata
${ M}_{\rm wd}$  &  $1.00{\pm}0.2{M}_{\odot}$	 & $r_{\rm wd}$      &   $0.00771R_{\odot}$\\
${M}_{\rm sec}$  &  $0.30{\pm}0.10{M}_{\odot}$	     & log $g_{\rm wd}$  &   8.35\\
${\dot{M}}$      &  $3.0{\pm}1.0{\times}10^{-9}{M}_{\odot} {\rm yr}^{-1}$ & $r_s$(pole) &  $0.311R_{\odot}$\\
P    &  0.133 days	  & $r_s$(point)   & $0.451R_{\odot}$\\                   
$D$              &  $1.1901R_{\odot}$   	 & $r_s$(side)  & $0.324R_{\odot}$\\
${\Omega}_{\rm wd}$         & 155.7       & $r_s$back)   & $0.363R_{\odot}$\\
${\Omega}_s$                &  2.46623        & log $g_s$(pole) & 4.95\\
{\it i}              &   $65{\pm}5{\degr}$	 & log $g_s$(point) & -0.44\\
$T_{\rm eff,wd}$         &  $45,000{\pm}5000$K       & log $g_s$(side)  & 4.88\\
$T_{\rm eff,s}$(pole)         &  3500K(nominal)  & log $g_s$(back)  & 4.67\\    
$A_{\rm wd}$                       &  1.0 & $r_a$ & $0.39R_{\odot}$\\      
$A_s$                       &  0.5       		 & $r_b$ & $0.00771R_{\odot}$\\
$b_{\rm wd}$                       &  0.25 & $H$    & $0.0096R_{\odot}$\\     
$b_s$                       &  0.08\\      
\enddata

\tablecomments{${\rm wd}$ refers to the WD; $s$ refers to the secondary star.
$D$ is the component separation of centers,
${\Omega}$ is a Roche potential. Temperatures are polar values, 
$A$ values are bolometric albedos, and $b$ values are 
gravity-darkening exponents. 
$r_a$ specifies the outer radius 
of the accretion disk, set at the tidal cut-off radius, 
and $r_b$ is the accretion disk inner radius, as 
determined in the final system model, while $H$ is 
the semi-height of the accretion disk rim, based on the standard model.}  
\end{deluxetable}

\begin{deluxetable}{cccc}
\tablewidth{0pt}
\tablenum{4}
\tablecaption{Features of SDSSJ0809 Balmer emission lines 
}
\tablehead{	\colhead{Identification} &  
\colhead{FWHM} & \colhead{EW}  & \colhead{decrement}  
}
\startdata
$H{\alpha}$   &   21.5  &  21.6  &	 2.5\\
$H{\beta}$    &   14.0  &	8.78 &   1.00\\
$H{\gamma}$   &   13.0  &	5.98 &   0.68\\
$H{\delta}$   &   12.0  &	4.85 &   0.55\\
$H{\epsilon}$ &   14.0\tablenotemark{a}  &	 \nodata & \nodata\\
$H8$		  &   11.0\tablenotemark{b}	&	 \nodata & \nodata\\
$H9$		  &    6.0	&	1.00 &   0.11\\
\enddata
\tablenotetext{a}{Blended with He II 3968.43\AA}
\tablenotetext{b}{Blended with He II 3887.44\AA}
\end{deluxetable}

\begin{deluxetable}{rrrrrr}
\tablewidth{0pt}
\tablenum{5}
\tablecaption{Accretion disk temperature profile for discovery SDSS spectrum}
\tablehead{
\colhead{$r/r_{\rm wd}$} & \colhead{$T_{\rm eff}$} & \colhead{$r/r_{\rm wd}$} & \colhead{$T_{\rm eff}$} & 
\colhead{$r/r_{\rm wd}$} & \colhead{$T_{\rm eff}$}
}
\startdata
1.0000 &     14,000    		& 16.3448 &		14,000			   & 34.6581 &	 9079\\
1.0225 &	 14,000			& 18.0096 &		14,000			   & 36.3230 &	 8776\\
1.3611 &	 14,000			& 19.6745 &		13,644			   & 37.9878 &	 8495\\
3.0260 &	 14,000			& 21.3393 &		12,875			   & 39.6527 &	 8234\\
4.6908 &	 14,000			& 23.0042 &		12,200			   & 41.3175 &	 7992\\
6.3557 &	 14,000			& 24.6690 &		11,164			   & 42.9824 &	 7766\\
8.0205 &	 14,000			& 26.3339 &		11,071			   & 44.6472 &	 7554\\
9.6854 &	 14,000			& 27.9987 &		10,593			   & 46.3121 &	 7355\\
11.3502 &	 14,000			& 29.6636 &		10,162			   & 47.9769 &	 7168\\
13.0151 &	 14,000			& 31.3284 &		9768			   & 49.6418 &	 6992\\
14.6799 &	 14,000			& 32.9933 &		9409			   & 51.3066 &	 6826\\
\enddata
\end{deluxetable}

\begin{deluxetable}{rrrrrr}
\tablewidth{0pt}
\tablenum{6}
\tablecaption{Accretion disk temperature profile for archival SDSS spectrum}
\tablehead{
\colhead{$r/r_{\rm wd}$} & \colhead{$T_{\rm eff}$} & \colhead{$r/r_{\rm wd}$} & \colhead{$T_{\rm eff}$} & 
\colhead{$r/r_{\rm wd}$} & \colhead{$T_{\rm eff}$}
}
\startdata
1.0000 &     11,500    		& 16.3448 &		10,000			   & 34.6581 &	 9000\\
1.0225 &	 11,500			& 18.0096 &		10,000			   & 36.3230 &	 8500\\
1.3611 &	 11,500			& 19.6745 &		10,000			   & 37.9878 &	 8300\\
3.0260 &	 11,500			& 21.3393 &		10,000			   & 39.6527 &	 8100\\
4.6908 &	 11,000			& 23.0042 &		10,000			   & 41.3175 &	 7800\\
6.3557 &	 11,000			& 24.6690 &		10,000			   & 42.9824 &	 7500\\
8.0205 &	 11,000			& 26.3339 &		10,000			   & 44.6472 &	 7200\\
9.6854 &	 11,000			& 27.9987 &		10,000			   & 46.3121 &	 7000\\
11.3502 &	 10,500			& 29.6636 &		10,000			   & 47.9769 &	 6700\\
13.0151 &	 10,500			& 31.3284 &		9500			   & 49.6418 &	 6500\\
14.6799 &	 10,500			& 32.9933 &		9200			   & 51.3066 &	 6200\\
\enddata
\end{deluxetable}

\begin{figure}[tb]
\epsscale{0.97}
\plotone{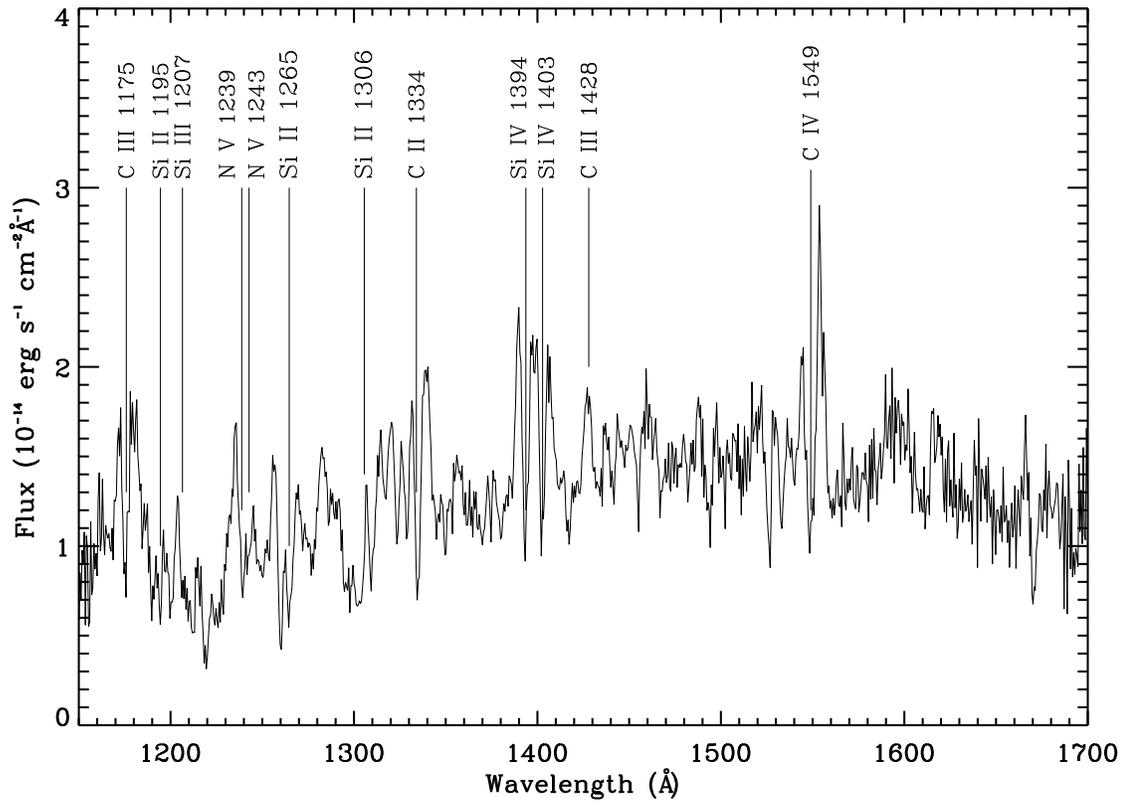}
\epsscale{1.00}
\figcaption{{\it HST} spectrum of SDSSJ0809. Note that several emission features have
absorption cores extending below the continuum, indicating a substantial
line-of-sight column density. 
\label{HST spectrum of SDSSJ0809}}
\end{figure}

\clearpage

\begin{figure}[tb]
\epsscale{0.97}
\plotone{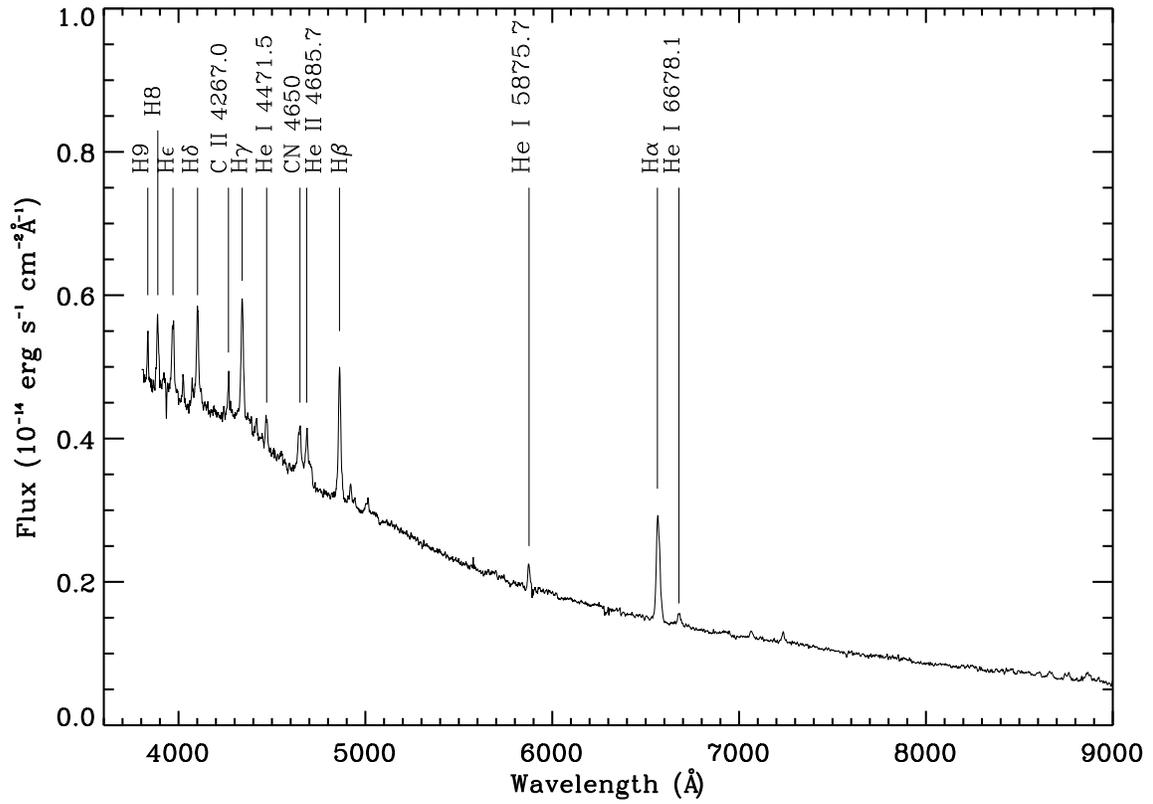}
\epsscale{1.00}
\figcaption{Discovery SDSS spectrum of SDSSJ0809. Note the lack of absorption wings
on the Balmer lines and the absence of any indication of a Balmer
discontinuity. 
\label{SDSS spectrum of SDSSJ0809}}
\end{figure}

\begin{figure}[tb]
\epsscale{0.97}
\plotone{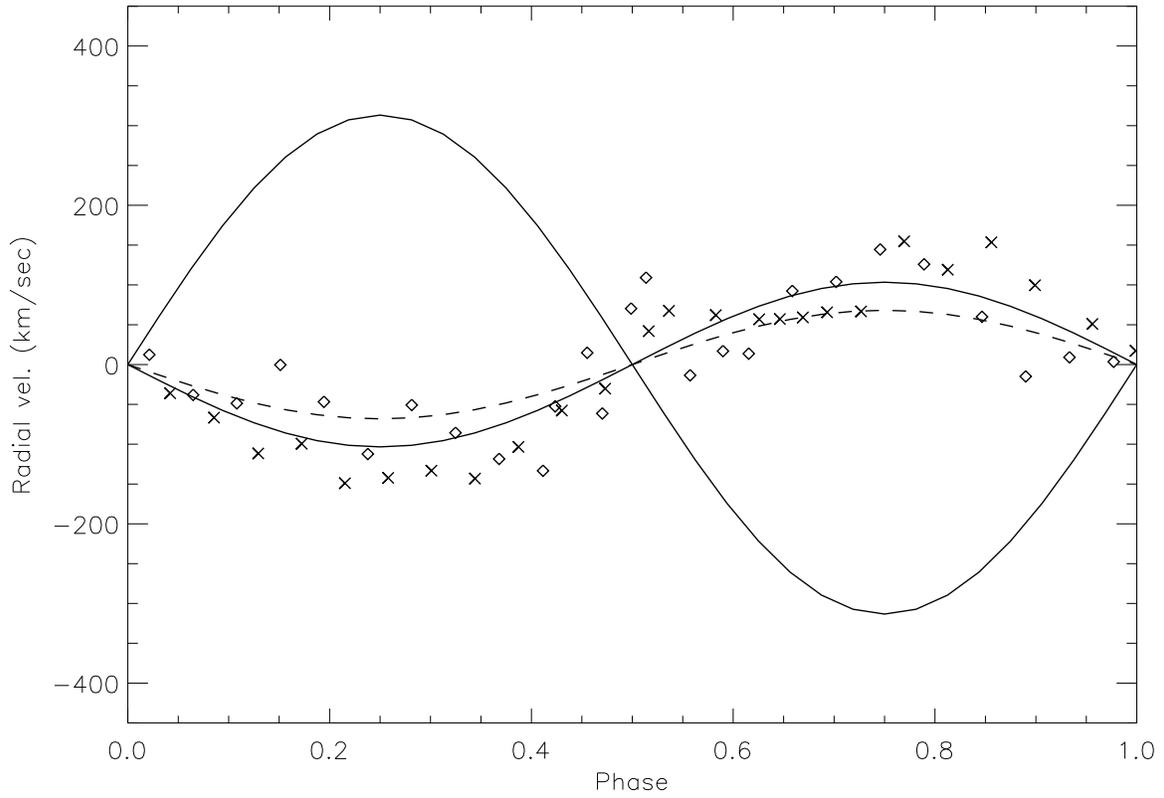}
\epsscale{1.00}
\figcaption{Theoretical radial velocity curves of system stars compared
with H$\alpha$ (diamonds) and H$\beta$ (x's) observations. The WD
mass is 1.0$M_{\odot}$, the mass ratio is $q=0.30$, the orbital period
is $192^{\rm m}$, and the orbital inclination is $65{\degr}$ (continuous curves). 
The dashed curve is the projected WD velocity for a mass ratio of $q=0.20$.
\label{radial velocity curve}}
\end{figure}

\begin{figure}[tb]
\epsscale{0.97}
\plotone{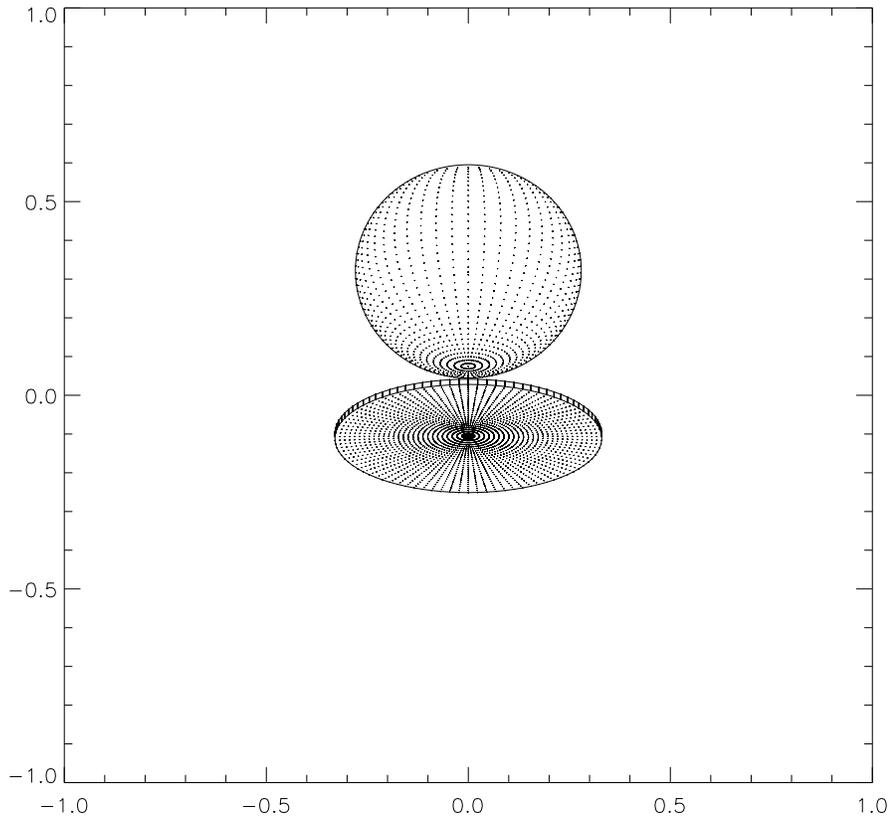}
\epsscale{1.00}
\figcaption{View of the system described in Figure~5 at orbital phase 0.0, $q=0.30$, 
projected on the plane of the sky for $i=65{\degr}$.
This is inferior conjunction for the secondary star; it is
between the observer and the WD.
The inner edge of the accretion disk extends to the WD equator
in this plot (no truncation), and the outer accretion disk radius is
at the tidal cutoff radius.	The plot for a $q=0.20$, $i=67{\degr}$ system is closely
similar.
\label{projection of system}}
\end{figure}

\begin{figure}[tb]
\epsscale{0.97}
\plotone{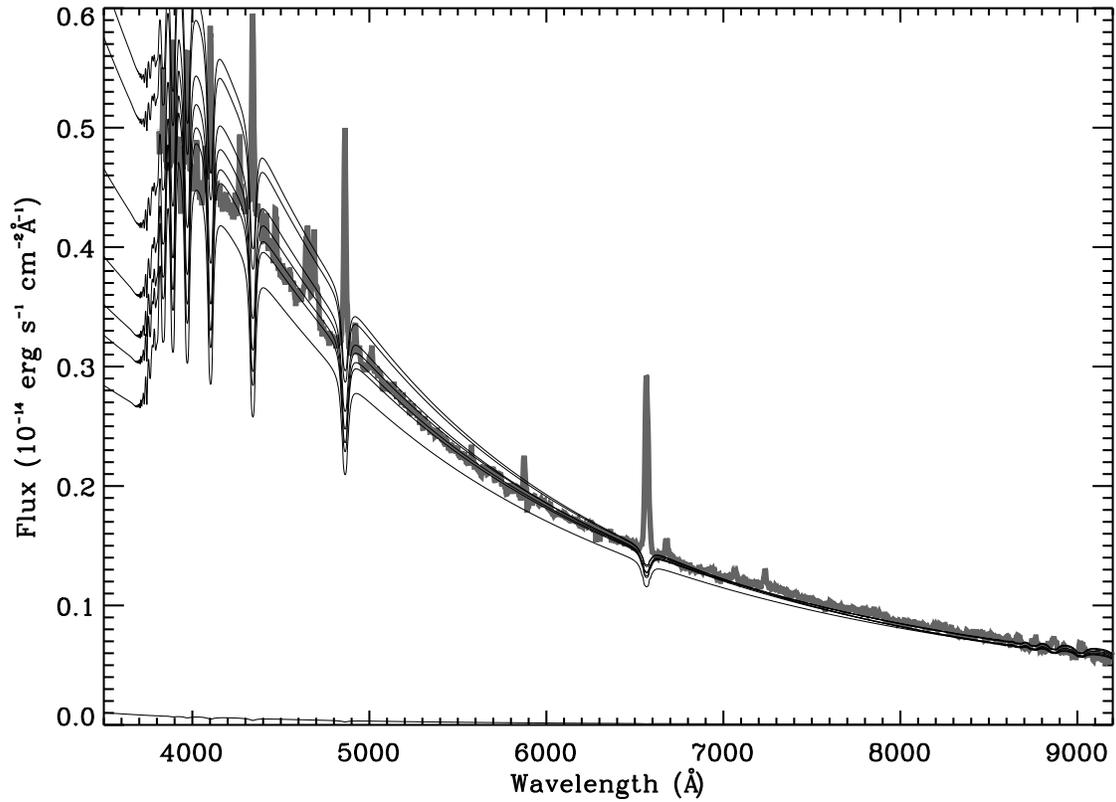}
\epsscale{1.00}
\figcaption{Synthetic spectra for truncation radii of (from top to bottom)
1.0, 4.0, 8.0, 12.0, 14.3, 16.0, and 18.0 
times the WD radius (thin lines), fitted
to the SDSS spectrum (broad line).
The mass transfer rate is $3.0{\times}10^{-9}M_{\odot}~{\rm yr^{-1}}$.
The best fit is the $r=16.0r_{WD}$ spectrum; it shows separately in Figure~10.
\label{3p0c}}
\end{figure}

\begin{figure}[tb]
\epsscale{0.97}
\plotone{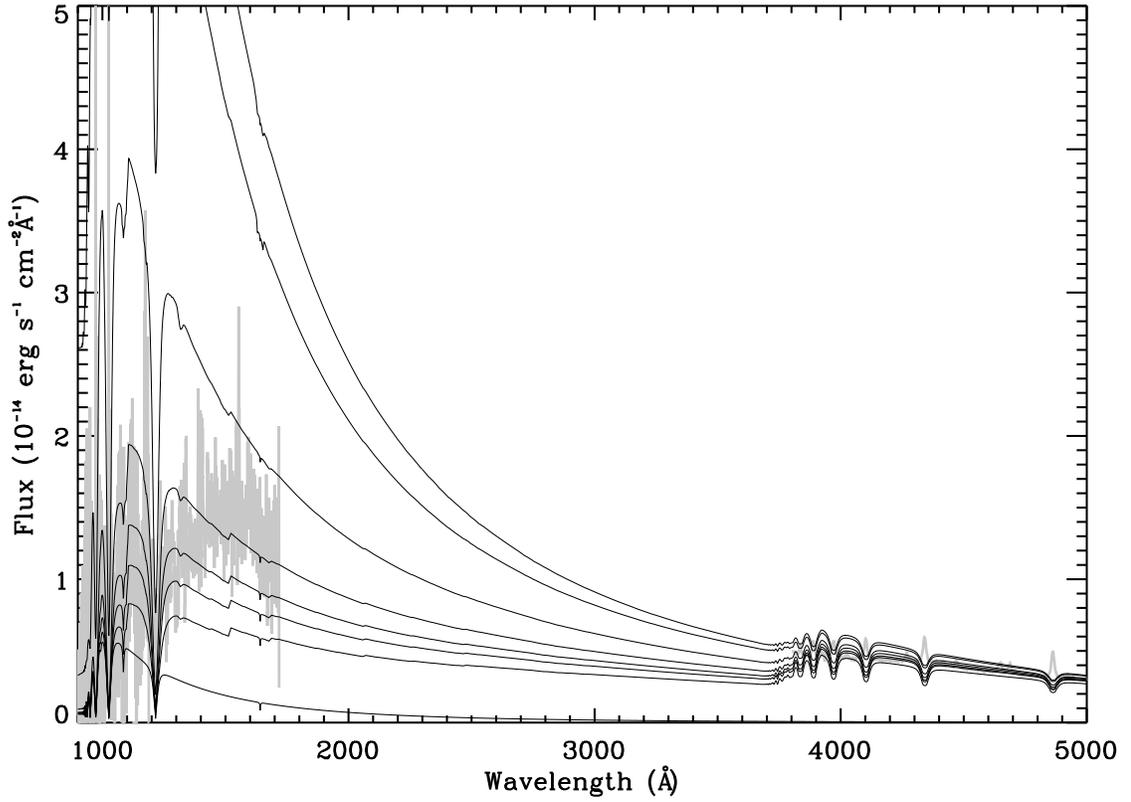}
\epsscale{1.00}
\figcaption{
As in Figure~7, but extending into the UV region.
Note the large separation of synthetic spectra in the UV for a relatively small
difference in the optical.
The mass transfer rate is $3.0{\times}10^{-9}M_{\odot}~{\rm yr^{-1}}$.
\label{3p0b}}
\end{figure}

\begin{figure}[tb]
\epsscale{0.97}
\plotone{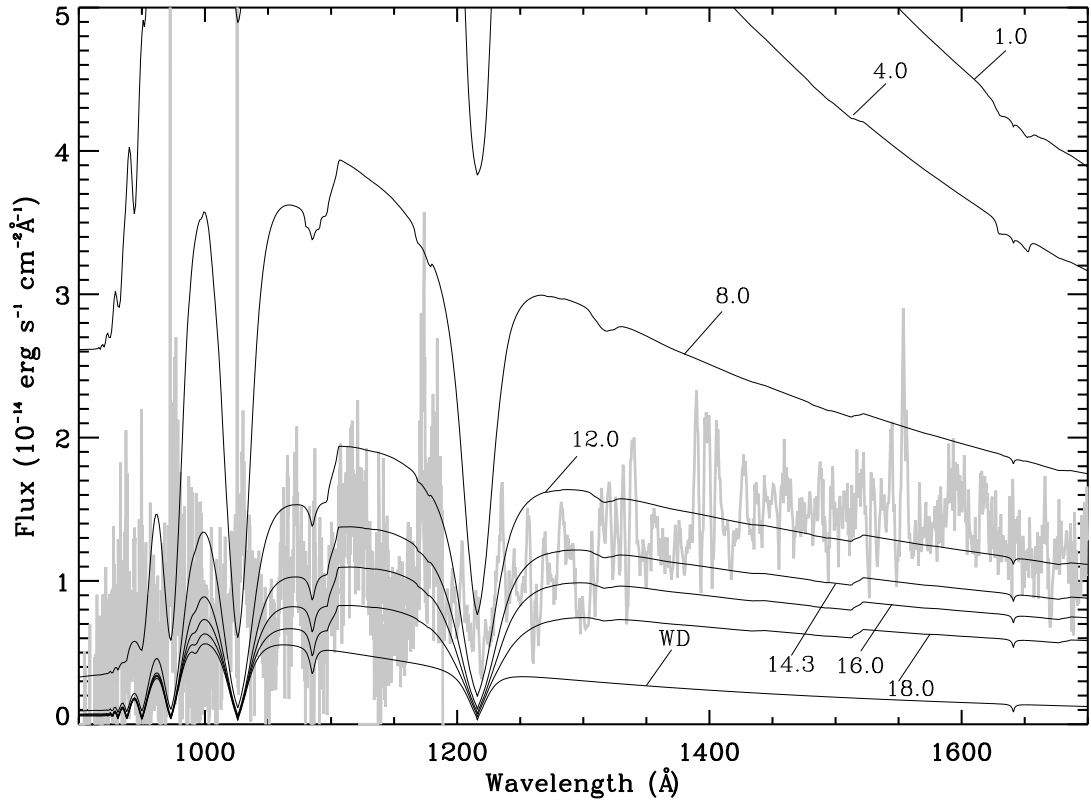}
\epsscale{1.00}
\figcaption{ As in Figure~8 but showing only the UV. 
The 35,000K WD model spectrum is at the bottom. The other model
spectra are labeled with the disk inner truncation radius (in WD radii).
\label{3p0a}}
\end{figure}

\begin{figure}[tb]
\epsscale{0.97}
\plotone{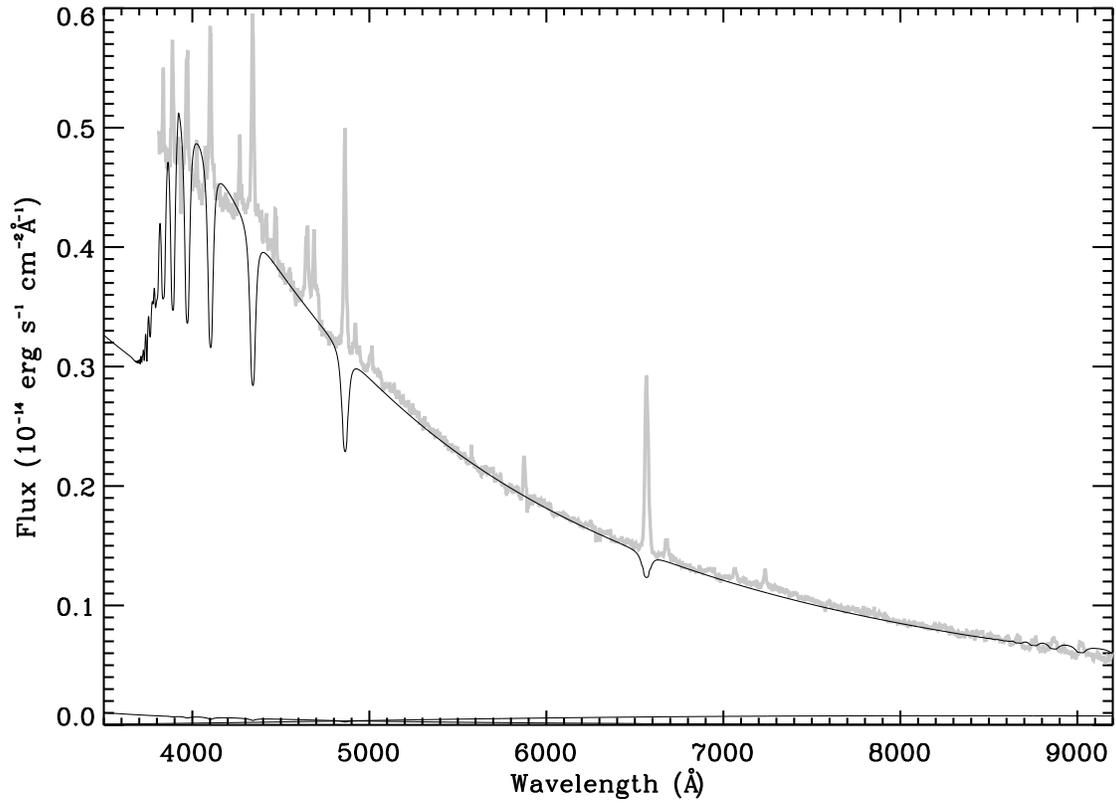}
\epsscale{1.00}
\figcaption{ 
Plot showing detail of the $r_{\rm inner}=16.0r_{WD}$ model(dark line) fit 
to the optical spectrum (light line).
The spectrum rising to the bottom right is the secondary star component, 
and the spectrum
rising to the bottom left is the 35,000K WD component.
\label{3p0d}}
\end{figure}

\begin{figure}[tb]
\epsscale{0.97}
\plotone{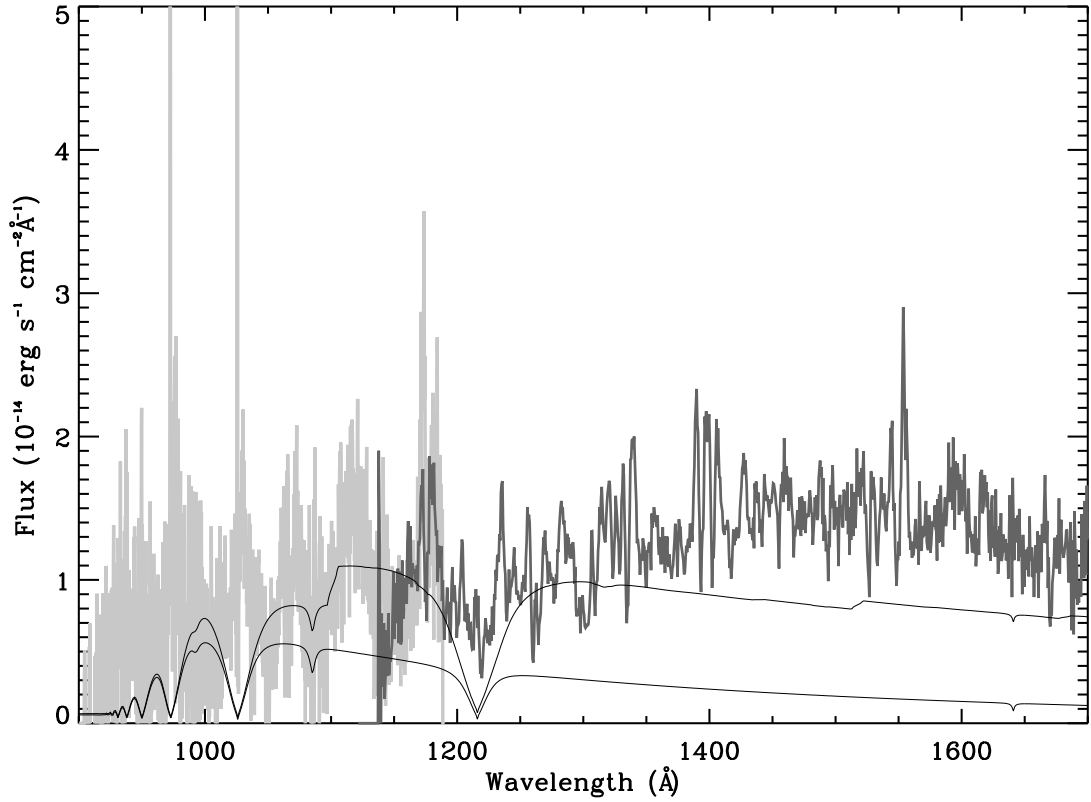}
\epsscale{1.00}
\figcaption{ As in Figure~10 but for the UV data. Note that the $FUSE1$
and $HST$ spectra agree well in the overlap region.
The 35,000K WD contribution is the lowest spectrum. 
\label{3p0i}}
\end{figure}

\begin{figure}[tb]
\epsscale{0.97}
\plotone{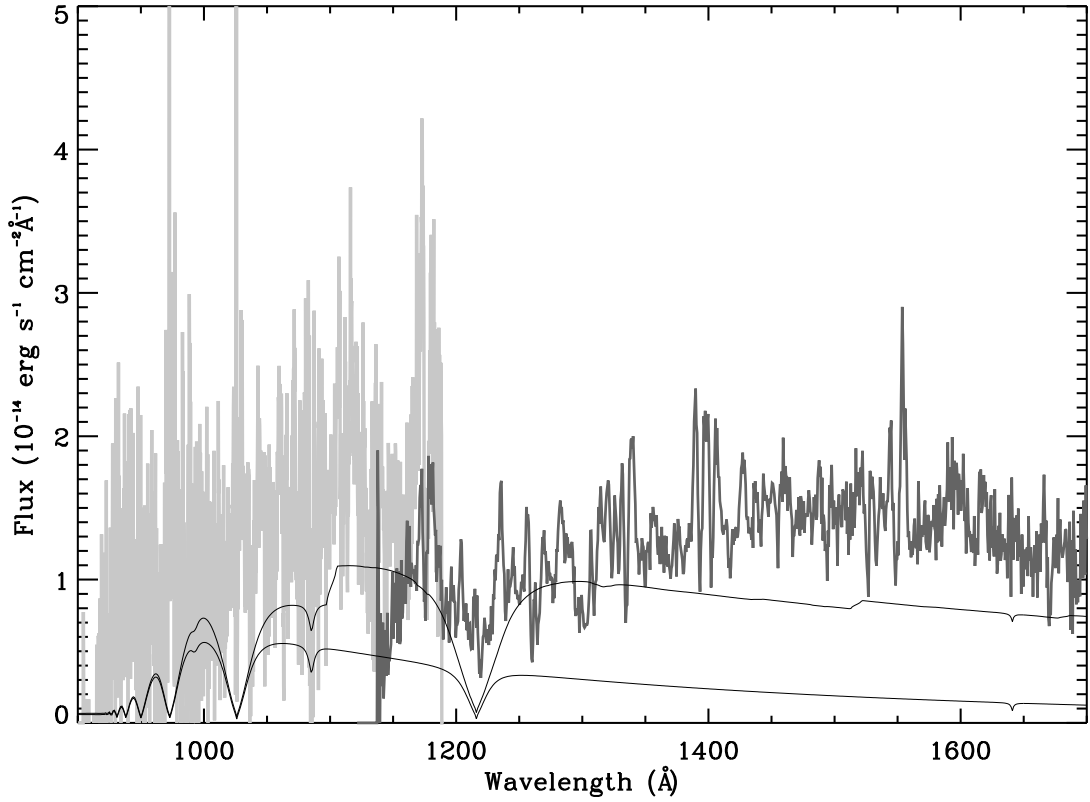}
\epsscale{1.00}
\figcaption{ 
As in Figure~11 except the $FUSE2$ spectrum has been substituted for the
$FUSE1$ spectrum. The $HST$ and $FUSE2$ spectra do not match well
in the overlap region.
The synthetic spectrum is a poor fit to the $FUSE2$ spectrum.
\label{3p0if2}}
\end{figure}

\begin{figure}[tb]
\epsscale{0.97}
\plotone{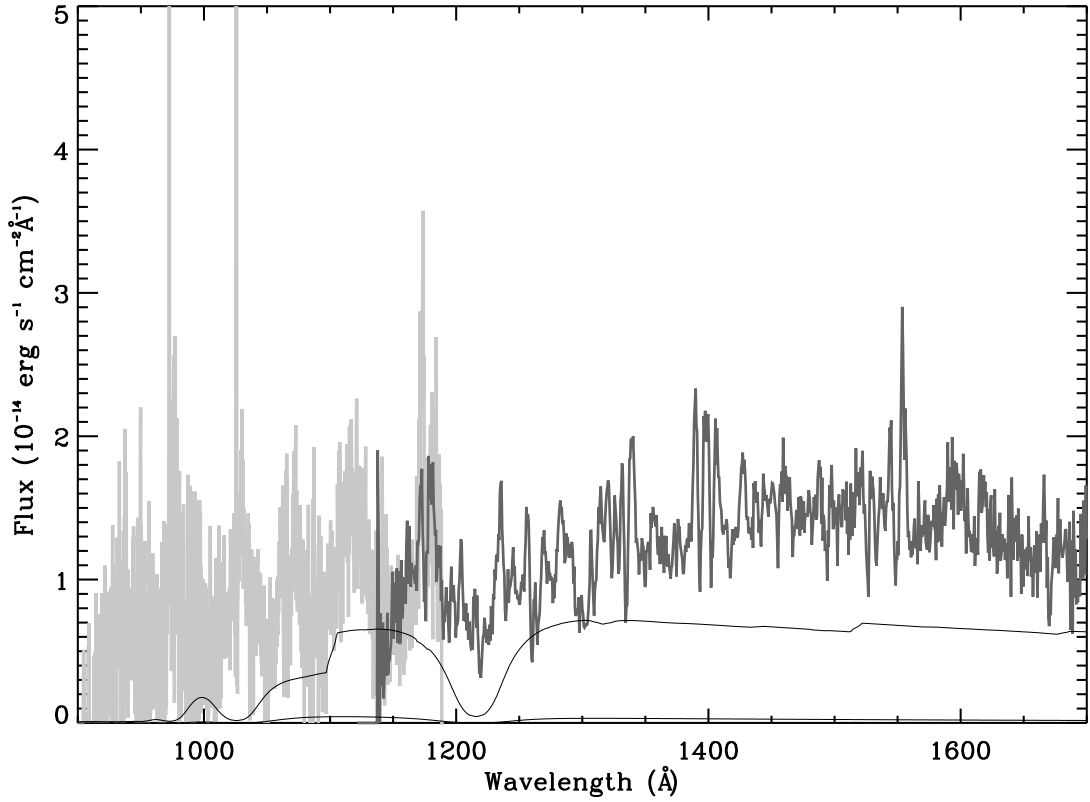}
\epsscale{1.00}
\figcaption{
As in Figure~11 except a 20,000K WD has been substituted.
The synthetic spectrum now passes through the middle of the $FUSE1$ 
spectrum near 1150\AA, but the flux deficit longward of 1300\AA~is appreciable.
\label{3p0e}}
\end{figure}

\clearpage

\begin{figure}[tb]
\epsscale{0.97}
\plotone{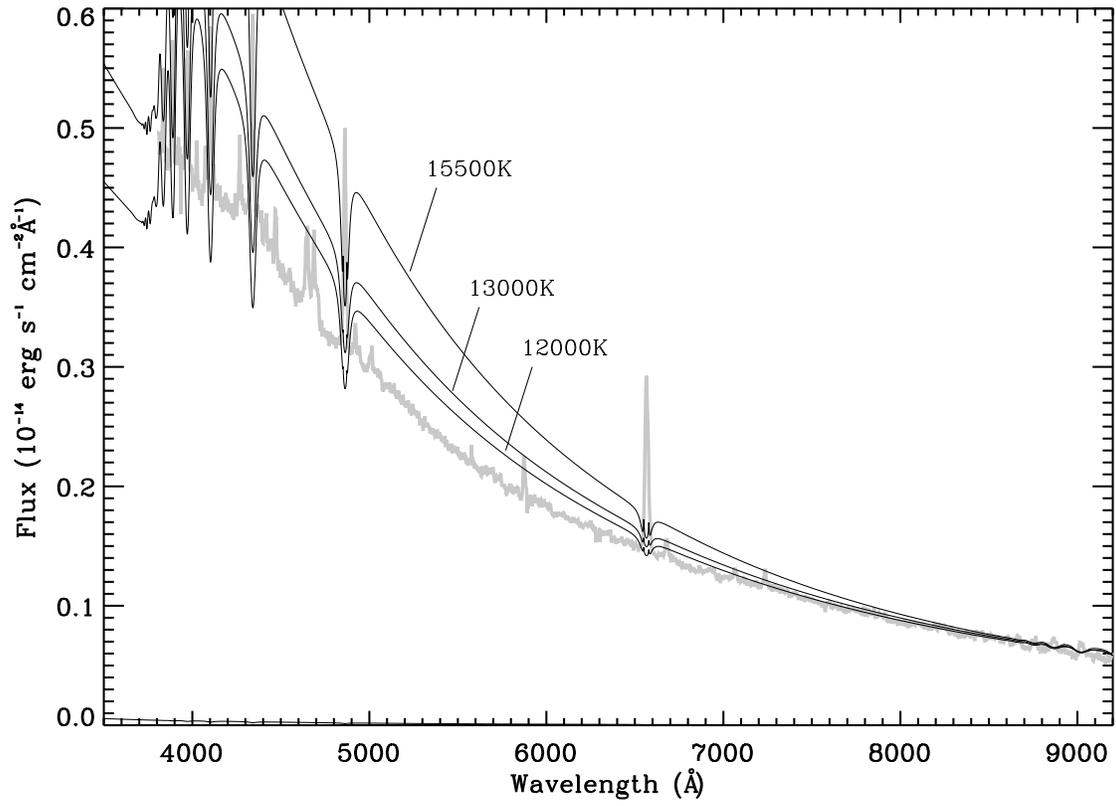}
\epsscale{1.00}
\figcaption{Synthetic spectrum for three isothermal accretion disks
(untruncated).
Note that the spectral gradient is too large in all cases.
\label{1p0isoa spectrum fit at long wavelengths}}
\end{figure}

\begin{figure}[tb]
\epsscale{0.97}
\plotone{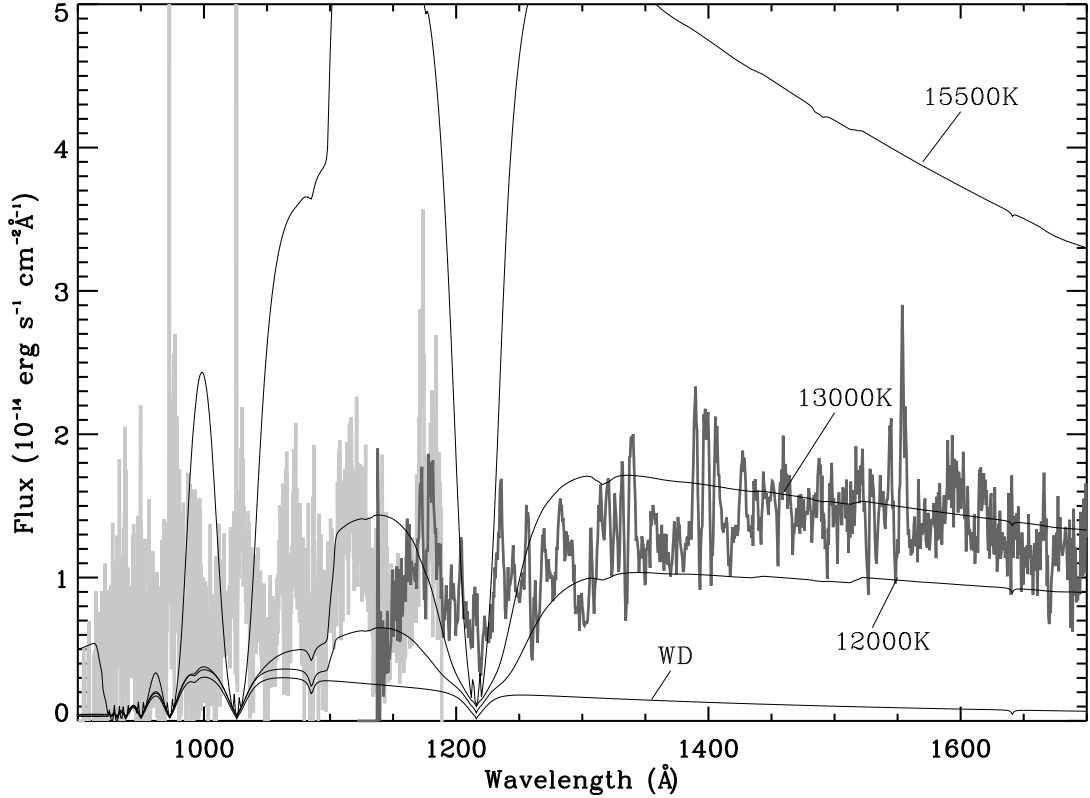}
\epsscale{1.00}
\figcaption{As in Figure~9 but showing the isothermal disk models from Figure~14.	
The relative contribution of the 35,000K WD
is less than in Figure~9 because the required scaling 
factor to superpose the spectra is larger in this case since the accretion
disk is not truncated. Note that there is a large discrepancy from the $FUSE1$ spectrum.
The 12,000K synthetic spectrum
provides the best fit to the $HST$ and $FUSE1$ spectra, but
the corresponding fit in the optical region is poor.
\label{1p0isob spectrum fit at long wavelengths}}
\end{figure}

\begin{figure}[tb]
\epsscale{0.97}
\plotone{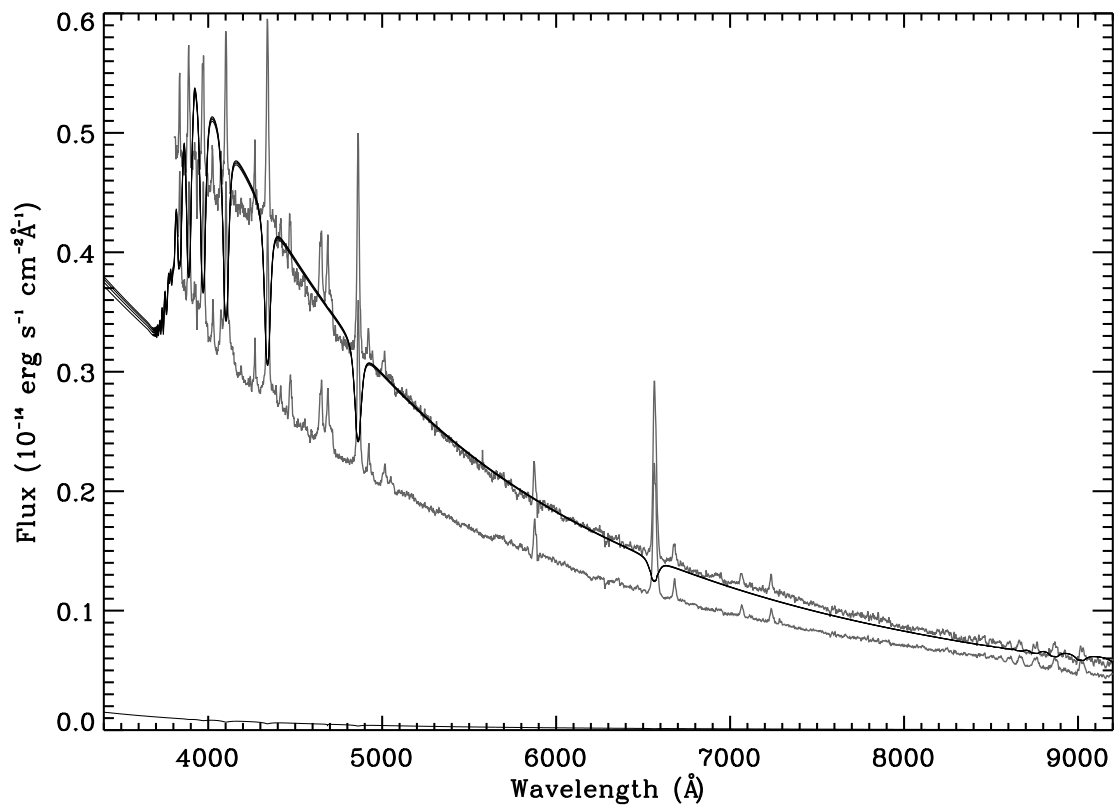}
\epsscale{1.00}
\figcaption{
Models with four different values of WD $T_{\rm eff}$ and an accretion disk that is
isothermal in the inner region and follows a standard model in the outer region.
The accretion disk temperature profile is in Table~5.
The models fit the discovery SDSSJ0809 spectrum. The other observed spectrum is
from the archive.
The models are insensitive to the
WD $T_{\rm eff}$ at optical wavelengths and are essentially indistinguishable
in the figure.
The 45,000K WD contribution is at the bottom.\label{3p0k}}
\end{figure}

\begin{figure}[tb]
\epsscale{0.97}
\plotone{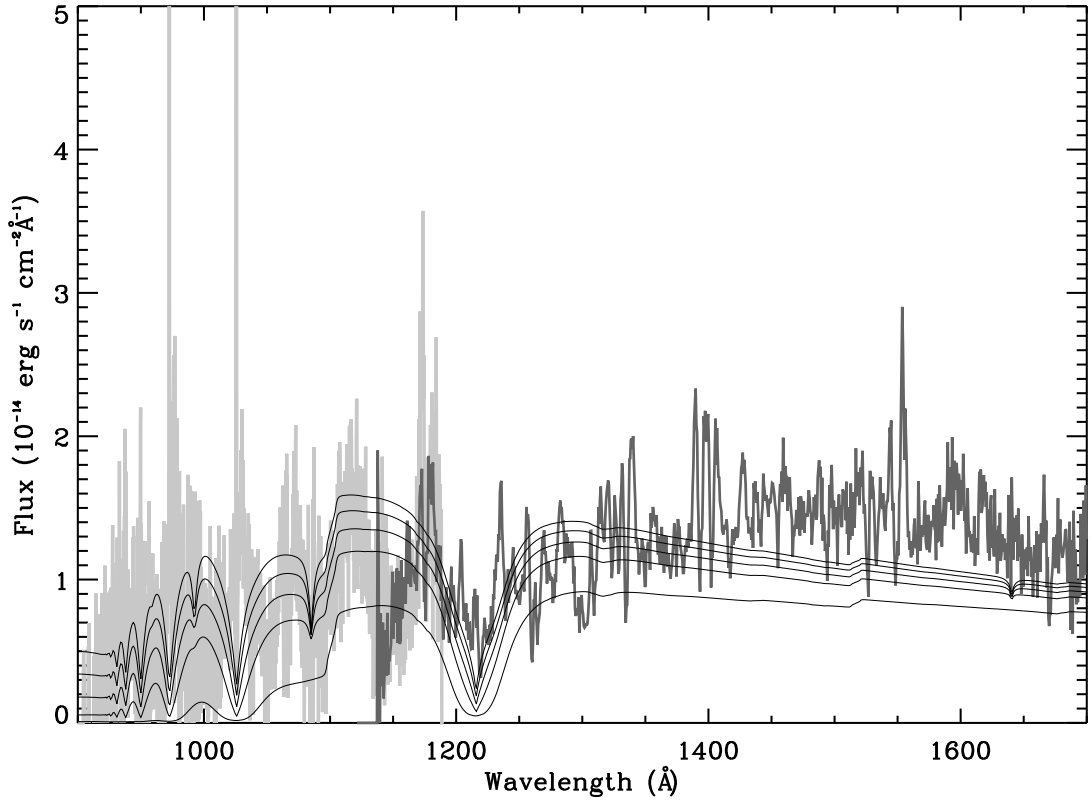}
\epsscale{1.00}
\figcaption{
As in Figure~16 except in the UV. From top to bottom, the synthetic spectra
show the models containing a WD with $T_{\rm eff}=$ 50,000K, 45,000K, 40,000K,
and 35,000K. The lowest synthetic spectrum is the contribution of
the accretion disk only.
\label{3p0m}}
\end{figure}

\begin{figure}[tb]
\epsscale{0.97}
\plotone{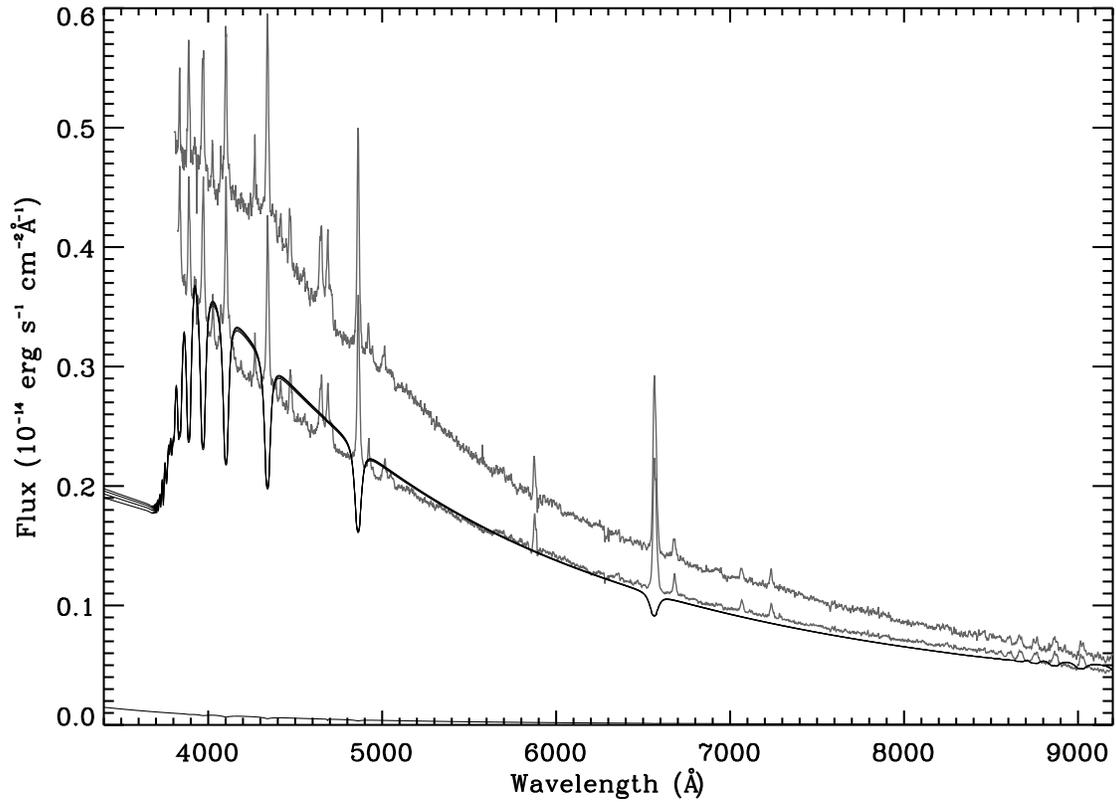}
\epsscale{1.00}
\figcaption{
Models with four different values of WD $T_{\rm eff}$ and an accretion disk 
whose temperature profile is in Table~6. The models fit the SDSSJ0809 spectrum
from the archive. Compare with Figure~16.
The lowest curve is the contribution of a 45,000K WD.
}
\end{figure}

\begin{figure}[tb]
\epsscale{0.97}
\plotone{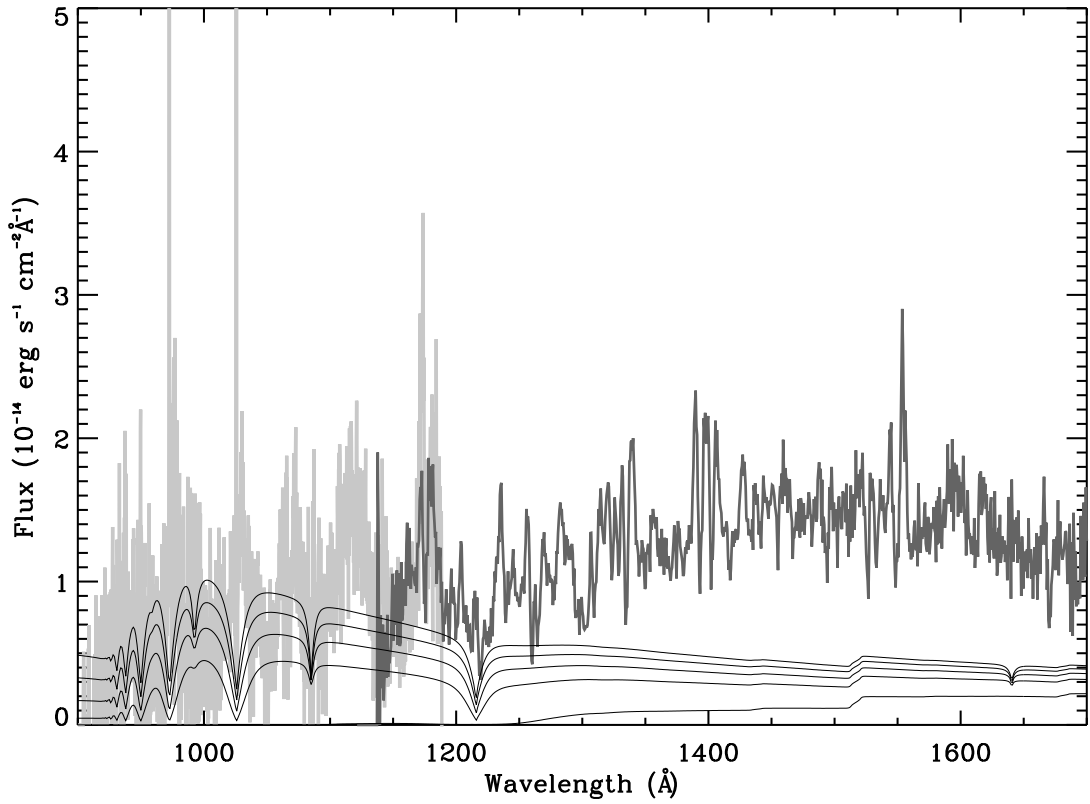}
\epsscale{1.00}
\figcaption{
As in Figure~18 but showing the UV fit. From top to bottom, the curves are
models containing a WD with $T_{\rm eff}=$
50,000K, 45,000K, 40,000K, and 35,000K.	See the text for a discussion.
The lowest synthetic spectrum is the accretion disk contribution only.
}
\end{figure}

\clearpage

\end{document}